\documentclass[submission,copyright,creativecommons]{eptcs}
\providecommand{\event}{LSFA 2011} 
\usepackage{breakurl}             

\providecommand{\urlalt}[2]{\href{#1}{#2}}
 \providecommand{\doi}[1]{doi:\urlalt{http://dx.doi.org/#1}{#1}}
\usepackage{graphicx}
\usepackage[all,knot]{xy}
\xyoption{arc} 

\usepackage{amsmath,amssymb,amsthm}
\usepackage{pslatex} 

\title{On Graph Refutation for Relational Inclusions\thanks{Research partly sponsored by the Brazilian  agencies CNPq and FAPERJ.}}
\author{Paulo A.~S.  Veloso
\institute{COPPE-UFRJ \\ Systems and Computer Engin. Program\\
UFRJ: Federal University of Rio de Janeiro\\
RJ, Brazil}
\email{pasveloso@gmail.com}
\and
\qquad\qquad Sheila R. M. Veloso 
\institute{FEN-UERJ \\ 
Systems and Computer Engin. Dept., Fac. of Engineering \\ 
  UERJ: State University of Rio de Janeiro\\
    RJ , Brazil}
\email{
 \quad\qquad sheila.murgel.bridge@gmail.com}
}
\def\titlerunning{Graph Refutation for Relational Inclusions}
\def\authorrunning{Veloso \& Veloso}

\newtheorem{Thrm}{Theorem} [section]   
\newtheorem{Lem}{Lemma} [section]   
\newtheorem{Prop}{Proposition} [section] 
\newtheorem{Cor}{Corollary} [section] 
\newtheorem{Rem}{Remark} [section] 
\newtheorem{Exmpl}{Example} [section] 

\newcommand{\sm}{\small} 
\newcommand{\fns}{\footnotesize} 
\newcommand{\Ds}{\displaystyle} 
\newcommand{\Ss}{\scriptstyle} 
\newcommand{\SSs}{\scriptscriptstyle} 
\newcommand{\hs}{\hspace} 
\newcommand{\vs}{\vspace} 
\newcommand{\svs}{5} 
\newcommand{\nvs}{6} 
\newcommand{\mvs}{8} 
\newcommand{\lvs}{10} 

\newcommand{\ul}[1]{\underline{{#1}}} 
\newcommand{\ol}[1]{\overline{{#1}}} 
\newcommand{\strl}[2]{\stackrel{#1}{#2}} 
\newcommand{\strld}[2]{\stackrel{{\Ds #1}}{#2}} 

\newcommand{\Lm}{\left} 
\newcommand{\Rm}{\right} 
\newcommand{\Ba}{\begin{array}} 
\newcommand{\Ea}{\end{array}} 
\newcommand{\Bt}{\begin{tabular}} 
\newcommand{\Et}{\end{tabular}} 

\newcommand{\Setof}[2]{ \{ {#1} \, \slash \, {#2} \} } 

\newcommand{\ES}{\emptyset} 

\newcommand{\Bi}{\begin{itemize}} 
\newcommand{\Ei}{\end{itemize}} 
\newcommand{\Be}{\begin{enumerate}} 
\newcommand{\Ee}{\end{enumerate}} 
\newcommand{\Bd}{\begin{description}} 
\newcommand{\Ed}{\end{description}} 

\newcommand{\Bop}{(} 
\newcommand{\Eop}{)} 
\newcommand{\BDr}{\Lag} 
\newcommand{\EDr}{\Rag} 
\newcommand{\BSl}{\Lag} 
\newcommand{\ESl}{\Rag} 
\newcommand{\drs}{,} 
\newcommand{\dps}{:} 
\newcommand{\ud}[1]{\underline{#1}} 

\newcommand{\BMd}{\Lag} 
\newcommand{\EMd}{\Rag} 
\newcommand{\BMr}{(} 
\newcommand{\EMr}{)} 

\newcommand{\rar}{\rightarrow}  
\newcommand{\lngrar}{\longrightarrow}  
\newcommand{\lar}{\leftarrow}  
\newcommand{\mpt}{\mapsto}  

\newcommand{\incl}{\subseteq}  

\newcommand{\Lag}{\langle} 
\newcommand{\Rag}{\rangle} 

\newcommand{\lbl}[1]{{\mathrm #1}} 
\newcommand{\snt}[1]{\mathsf{#1}} 
\newcommand{\nd}[1]{{\rm {#1}}} 
\newcommand{\sem}[1]{{\mathfrak{#1}}} 
\newcommand{\Rl}[1]{{\tt #1}} 
\newcommand{\cmp}{\cdot} 

\newcommand{\Rp}{\lbl{p}} 
\newcommand{\Rq}{\lbl{q}} 
\newcommand{\Rr}{\lbl{r}} 
\newcommand{\Rs}{\lbl{s}} 
\newcommand{\Rt}{\lbl{t}} 

\newcommand{\Ra}{\lbl{a}} 
\newcommand{\Rb}{\lbl{b}} 
\newcommand{\Rc}{\lbl{c}} 
\newcommand{\Rd}{\lbl{d}} 
\newcommand{\Rle}{\lbl{e}} 
\newcommand{\Rf}{\lbl{f}} 
\newcommand{\Rg}{\lbl{g}} 

\newcommand{\RP}{\lbl{P}} 
\newcommand{\RQ}{\lbl{Q}} 
\newcommand{\RR}{\lbl{R}} 

\newcommand{\RN}{\lbl{R\!n}} 

\newcommand{\ccmpl}{\widetilde{\hs{1pt}} } 
\newcommand{\trnsp}[1]{{#1}^{\tt T}} 
\newcommand{\csu}{\cup} 
\newcommand{\csa}{\cap} 

\newcommand{\opfnt}[1]{\mathsf{#1}} 

\newcommand{\btm}{\opfnt{I\!\!\!\!\bot}} 
\newcommand{\tp}{\opfnt{I\!\!\!\!\top}} 
\newcommand{\id}{\opfnt{I\!\!I}} 
\newcommand{\dv}{\opfnt{I\!\!D}} 
\newcommand{\cmpl}[1]{\ol{#1}} 
\newcommand{\cnv}[1]{{#1}^{{\SSs \opfnt{\smile}}}} 
\newcommand{\su}{\opfnt{\sqcup}} 
\newcommand{\sa}{\opfnt{\sqcap}} 
\newcommand{\rp}{\opfnt{;}} 
\newcommand{\rs}{\opfnt{\dagger}} 

\newcommand{\gop}{\bullet} 

\newcommand{\sqeq}{\sqsubseteq} 
\newcommand{\eq}{\equiv} 

\newcommand{\syn}[2]{{#1} \, \rp \,  {#2}}

\newcommand{\IN}{\mathsf{I\!Nd}} 

\newcommand{\x}{\nd{x}} 
\newcommand{\y}{\nd{y}} 
\newcommand{\z}{\nd{z}} 
\newcommand{\ndu}{\nd{u}} 
\newcommand{\ndv}{\nd{v}} 
\newcommand{\ndw}{\nd{w}} 

\newcommand{\dg}{\Sigma} 
\newcommand{\SlS}{\snt{S}} 
\newcommand{\SlT}{\snt{T}} 

\newcommand{\SlG}[1]{\snt{Sl}[#1]} 
\newcommand{\GrS}[1]{\snt{Gr}(#1)} 

\newcommand{\ada}{+} 
\newcommand{\ds}[2]{\snt{DS}( {#1} \setminus {#2})} 
\newcommand{\CS}[2]{\snt{Sl}(  #1 {\Ss \rar} #2 )} 
\newcommand{\gl}[4]{{#1} \frac{#2}{#3}{#4}} 

\newcommand{\tra}[1]{{#1}^{{\tt tr}}} 

\newcommand{\FRS}{\cal R} 

\newcommand{\SSH}{\Gamma} 

\newcommand{\mor}{\dasharrow} 
\newcommand{\Mor}[2]{\mathsf{Mor}[ {#1},{#2} ]} 

\newcommand{\ha}{\alpha} 
\newcommand{\hb}{\beta} 
\newcommand{\hth}{\theta} 
\newcommand{\hsg}{\sigma} 
\newcommand{\hta}{\tau} 

\newcommand{\red}{\rhd} 
\newcommand{\cnvu}[1]{\strl{(#1)}{\red}} 
\newcommand{\clrd}{\red^{\ast}} 

\newcommand{\ex}{\lhd} 
\newcommand{\clrex}{\ex^{\ast}} 

\newcommand{\rd}[1]{{#1}^{\tt bs}} 

\newcommand{\Der}{\vdash} 

\newcommand{\Exp}{\Rl{Exp}} 

\newcommand{\Hyp}[1]{\Rl{Hyp}[{#1}]} 

\newcommand{\dc}{\cmpl{\cmpl{\hs{4pt}}}} 

\newcommand{\ISS}[1]{\SLI[{#1}]} 

\newcommand{\gM}{\sem{M}} 
\newcommand{\gC}{\sem{C}} 

\newcommand{\g}{{\tt g}}  
\newcommand{\rel}[1]{{{\mathsf[}}#1{{\mathsf ]}}}  
\newcommand{\arcsat}[1]{\Vdash_{#1}} 

\newcommand{\betM}[1]{{[\![}#1{]\!]_{\gM}}} 
\newcommand{\betCn}[1]{{[\![}#1{]\!]_{\gC}}} 

\newcommand{\SLI}{\Lambda} 

\newcommand{\ESL}[1]{{\mathsf{ES}}[#1]}  
\newcommand{\ESA}[1]{{\mathsf{ES}}[#1]}  
\newcommand{\ESD}[1]{{\tt ES}[#1]}  
\newcommand{\ESS}[1]{{\tt ES}[#1]}  
\newcommand{\rkd}[1]{{\tt rk}(#1)}  
\newcommand{\rka}[1]{{\mathsf{rk}}(#1)}  
\newcommand{\rkl}[1]{{\mathsf{rk}}(#1)}  
\newcommand{\rks}[1]{{\tt rk}(#1)}  

\newcommand{\aro}[1]{\ar^{{\Ds #1}}} 
\newcommand{\aru}[1]{\ar_{{\Ds #1}}} 

\newcommand{\inp}{{\Ss \rightarrow}} 
\newcommand{\out}{{\Ss \leftarrow}} 
\newcommand{\dinp}{{\Ss \downarrow}} 
\newcommand{\upout}{{\Ss \uparrow}} 
\newcommand{\dsz}{2}  

\begin{document}
\maketitle

\begin{abstract}
We introduce a   graphical refutation calculus for relational inclusions:  
it reduces establishing a relational inclusion to establishing that a graph constructed from it has empty extension. 
This sound and complete calculus is conceptually simpler and easier to use than the usual ones.

\end{abstract}

\section{Introduction} \label{sec:Intr} 

We introduce a sound and complete goal-oriented graph calculus for relational inclusions.\footnote{Discussions 
with Petrucio Viana and Renata de Freitas are gratefully acknowledged.}  
Though somewhat richer, 
it is conceptually simpler and easier to use than the usual ones, 
as is its extension for handling hypotheses, due to goal-orientation. 

	Diagrams and figures are very important and useful  in several branches of science, as well as in everyday life.  
Graphs and diagrams provide convenient visualization in many 
areas~\cite{Bar_10,BH_94,BJ_94,Mad_96,ST_93}.  The heuristic appeal of diagrams is evident. 
Venn diagrams, for instance,  may be very helpful in visualizing connections between sets. 
They are not, however, usually accepted as proofs: one has to embellish the connections discovered in terms of 
standard methods of reasoning. This is not the case with our graph calculi: 
there is no need to compile the steps into standard reasoning. 
Graph manipulations, 
provided with precise syntax and semantics, are proof methods. 
 
		Formulas are usually written down on a single line~\cite{CL_96}.    
While  the   Polish parenthesis-free notation is more economical, the  usual notation  
is more readable:  
e.g. compare  $\rar \land p q \lor r s$ and $(p \land q) \rar (r \lor s)$. 
A basic idea behind graph calculi is a two-dimensional representation: e.g.  
the structure of $(x + y ) \cmp (z - w )$ is more apparent  in the 
notation 
$\Lm ( \Ba{c} x \\ + \\ y \Ea  \Rm ) \cmp  \Lm ( \Ba{c} z \\ - \\ w \Ea  \Rm )$  
(see also~\cite{Bar_10}). 
Using (individual) nodes in graph calculi is crucial, as well 
(see Sections~\ref{sec:BsIds} and~\ref{sec:GrLng}).

	Using drawings for relations is a  
	natural idea: represent the fact that $a$ is related to $b$ via 
relation $r$ by an arrow $a \, \strl{\Ds r}{\rar} \, b$. Then, some operations on relations correspond to simple manipulations on arrows, e.g. transposal 
to arrow reversal, intersection to parallel arcs and relative product to consecutive arcs  
(see Example~\ref{Exmpl:RAaxm}). 
So, one can reason about relations by manipulating their representations. 
This is a key idea 
underlying graph methods for reasoning about 
relations~\cite{CL_95,CL_96,FVVV_06,FVVV_07,FVVV_08,FVVV_09,FVVV_09',FVVV_10}. 
Some relational operations (like complementation) are not so easy to handle.\footnote{Complementation may be introduced by definition, if one can reason from 
hypotheses~\cite{FVVV_08}, or it can be handled    
via arcs labeled by boxes~\cite{FVVV_10}.}   
In this paper, we intend to show that one can profit from complementation 
by proposing a refutational graph calculus for reasoning about relations: this goal-orientated calculus, 
having  simpler  concepts, is easier to use than  the usual ones. 

The structure of this paper is as follows. 
In Section~\ref{sec:BsIds},  we illustrate the ideas underlying our calculus for relational inclusions.  
In Section~\ref{sec:GrLng}, we examine  our  graph language: syntax,  semantics as well as some concepts and constructions. 
In Section~\ref{sec:GrClc}, we introduce  our refutation calculus and its rules, 
which we extend to handle inclusion hypotheses in  Section~\ref{sec:Hyp}. 
Finally, Section~\ref{sec:Cncl} presents some remarks about our approach and 
other relational calculi. 


\section{Motivation: underlying ideas} \label{sec:BsIds} 
We now examine some basic ideas underlying our calculus for relational inclusions. 

	We wish to establish inclusions between relational terms. 
Relational terms are expressions like 
$\Rr$, $\cmpl{\Rs}$, $\Rr \, \sa \, \cmpl{\Rs}$ and $\cnv{\Rr} \, \rp \, \cmpl{(\Rr \, \rp \,  \Rs)}$. 
The  relational terms are (freely) generated from relation names 
by relational constants and operations,  as usual~\cite{Mad_06}. 
We employ the RelMiCs notation~\cite{BKS_77}. 
\Bi 
\item 
A  relation name $\Rr, \Rs, \Rt, \dots$ corresponds to an arbitrary binary relation (over a set $M$). 
\item 
The constants  \, $\btm$, $\tp$, $\id$ and $\dv$ denote respectively the following $2$-ary relations: 
 empty $\ES$, 
 square $M^2 := M \times M$,  identity   $I_M := \Setof{\Bop  a , b \Eop  \in M^2}{a = b}$ and 
 diversity $I_M \,  \ccmpl{} :=\Setof{\Bop  a , b \Eop  \in M^2}{a \neq b}$. 
\item 
The unary operations ${}^{\cmpl{\hs{5pt}}}$ and  $\cnv{}$ stand  for  Boolean complementation~$\, \ccmpl$ and Peircean transposition $\, \trnsp{}$. 
Recall that   
$\RR \,  \ccmpl{} := \Setof{\Bop  a , b \Eop \in M^2}{\Bop  a , b \Eop \not \in \RR}$ and 
 $\trnsp{\RR} := \Setof{\Bop  a , b \Eop \in M^2}{\Bop  b , a \Eop \in \RR}$.
\item  
The binary operations $\sa$ and  $\su$ stand for Boolean 
intersection $\csa$ and union $\csu$, respectively.
The binary operations  $\rp$ and  $\rs$  stand  for  relative product $\mid$ and sum $\ul{{\Ss |}}$, respectively. For $\Bop a, b \Eop \in M^2$, we have: 
$\Bop a, b \Eop \in \RP \mid \RQ$ iff, for some $c \in M$,  
	$\Bop a, c \Eop \in \RP$ and $\Bop c, b \Eop \in \RQ$, and 
$\Bop a, b \Eop \in \RP \, \ul{{\Ss |}} \, \RQ$ iff, for every $c \in M$,  
	$\Bop a, c \Eop \in \RP$ or $\Bop c, b \Eop \in \RQ$.  
\Ei 

	We can now 
	introduce the ideas of our graph methods 
(see also Sections~\ref{sec:GrLng} and~\ref{sec:GrClc}). 
	A graph is a finite set of alternative slices. 
	A slice consists of  finite sets of nodes and labeled arcs together 
with $2$ distinguished nodes (marked $\inp$). 
To establish an inclusion  $\RP \, \sqeq \,  \RQ$ we start with the slice corresponding to 
$\RP \, \sa \, \cmpl{\RQ}$ and apply the rules so as to  obtain 
a graph whose slices are inconsistent.  
	
	We now examine some simple examples illustrating our graph methods 
(see also Sections~\ref{sec:GrLng} and~\ref{sec:GrClc}). 

\begin{Exmpl}  \label{Exmpl:RAaxm} 
To establish $\cnv{\RP} \rp \, \cmpl{\RP \rp \RQ} \, \sqeq \,  \cmpl{\RQ}$, 
we show $( \cnv{\RP} \rp \, \cmpl{\RP \rp \RQ} )  \sa  \cmpl{\cmpl{\RQ}}  \sqeq \, \btm$. 

\Be  
\item 
First, we form a slice for $( \cnv{\RP} \rp \, \cmpl{\RP \rp \RQ} )  \sa  \cmpl{\cmpl{\RQ}}$,  with parallel arcs: $\SlS_0  := \xy
(0,0)*+{ \inp \, \x}="xS";
(30,0)*+{ \y \, \inp}="yS2";
{\aro{\cnv{\RP} \rp \, \cmpl{\RP \rp \RQ} }@/_-1pc/"xS";"yS2"};
{\aru{{ \cmpl{\cmpl{\RQ}}}}@/^-1pc/"xS";"yS2"}
\endxy$. 

\item We now convert this slice $\snt{S}_0$ to a special form, as follows.  
	\Be 
	\item We eliminate double complementation,  converting $\snt{S}_0$ to $\snt{S}_1$ as 
	follows: 
	
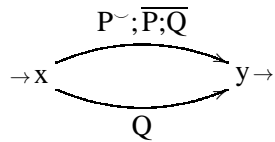
\[ 
	\xy
(0,0)*+{ \inp \, \x}="xS";
(30,0)*+{ \y \, \inp}="yS2";
{\aro{\cnv{\RP} \rp \, \cmpl{\RP \rp \RQ} }@/_-1pc/"xS";"yS2"};
{\aru{{\RQ}}@/^-1pc/"xS";"yS2"}
\endxy   	 \] 
	\item Next, we  eliminate $\rp$  by converting $\snt{S}_1$ to a $3$-node slice $\snt{S}_2$ as follows:  
	
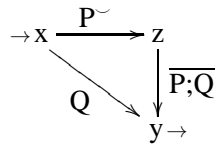
\[  \xymatrix{
	\inp \, \x \ar[r]^{{\Ds \, \cnv{\RP} }}  \ar[dr]_{{\Ds \RQ}} & \z \ar[d]^{ {\Ds \, \cmpl{\RP \rp \RQ} }} 
	  \\ & \hs{8 pt}   \y  \, \inp} \]  	 

	\item We eliminate $\cnv{}$ from $\snt{S}_2$, inverting its  
	arrow and giving $\snt{S}_3$ as follows: 
	
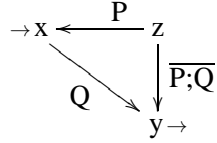
\[  \xymatrix{
	\inp \, \x \ar[dr]_{{\Ds \RQ}} & \z \ar[d]^{ {\Ds \, \cmpl{\RP \rp \RQ} }} \ar[l]_(.3){{\Ds \RP }} 
	  \\ & \hs{8 pt}   \y  \, \inp} \]  
	\item We now   convert $\snt{S}_3$ to $\snt{S}_4$ 
	(with a complemented slice as arc label): 
	
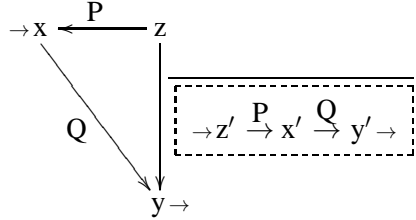
\[  \xymatrix{ 
	\inp \, \x \, \ar[ddr]_{{\Ds \RQ}} & \z \ar[dd] ^{\cmpl{ \mbox{ \dashbox{\dsz}(90, 26)[]{$\inp \, \z' \, 
	\strld{\RP}{\rar}  \x' \, 
	 \strld{ \RQ}{\rar} \, \y' \, \inp $}} }  }
	\ar[l]_{{\Ds \RP }} 
	  \\ 
	  \\
	& \hs{8 pt} \y  \, \inp}
	 \] 

	\Ee 
\item Now, within slice  $\snt{S}_4$, we have the following (parallel) paths from $\z$ to $\y$: 
	\Bi 
	\item 
positive path $\z \, \strl{{\Ds \RP}}{\rar} \, \x \,  \strl{{\Ds \RQ}}{\rar} \, \y$ 
(corresponding to the term $\RP \rp \RQ$) and 
	\item 
negative path 
 \xymatrix{
 \z \ar[rrrr] ^{\cmpl{ \mbox{ \dashbox{\dsz}(90, 26)[]{$\inp \z' \, \strld{ \RP}{\rar}  \x' \, 
	 \strld{ \RQ}{\rar} \, \y'  \, \inp $}} }  }
	 & & & &  \y  } (corresponding to the term $\cmpl{\RP \rp \RQ}$). 	 
	\Ei 
Slice $\snt{S}_4$ represents an inconsistent situation, 
corresponding to the empty relation $\ES$. 
\Ee 

\end{Exmpl}  

\begin{Exmpl}  \label{Exmpl:Mdlw} 
Consider the modular law $\RP \, \sqeq \,  \RQ$, where 
$\RP  :=  \Rr  \sa  (\Rs \rp \Rt) $ and 
$\RQ  := \Rs \, \rp [  (\cnv{\Rs} \rp \Rr )  \sa  \Rt ]$  (cf.~\cite{FVVV_09'}).  
We reduce it to 
$\RP \, \sa \, \cmpl{\RQ} \, \sqeq \, \btm$,  
which we can establish as follows (see Sections~\ref{sec:GrLng} and~\ref{sec:GrClc} for some details). 
\Be 
\item As before, we construct a  slice with parallel arcs, namely  
$\SlS := \xy
(0,0)*+{ \inp \, \x \, }="x"; 
(16,0)*+{ \, \y \, \inp}="y";
{\aro{\RP}@/_-1pc/"x";"y"};
{\aru{\cmpl{\RQ}}@/^-1pc/"x";"y"}
\endxy$.  

\item We can convert slice $ \snt{S}$ (see~\ref{subsec:Sntx} and~\ref{subsec:Redrls}) to an equivalent  slice $\snt{S}'$ with  slice $\cmpl{\snt{T}'}$ as arc label, where:  
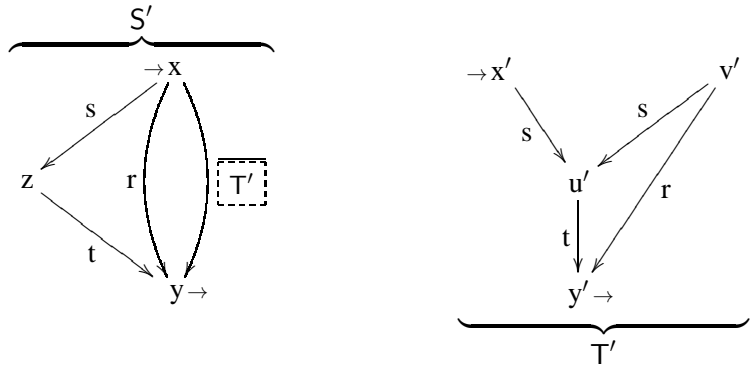
\[ \Ba{ccc}  
	\overbrace{ \xy
	(70,0)*+{ \inp \, \x \hs{10 pt}}="xS";
	(50,-15)*+{ \, \z}="uS"; 
	(70,-30)*+{ \hs{10 pt}  \nd{y}  \, \inp}="yS";
	\ar_{{\Ds \Rs}}"uS";"xS";
	\ar_{{\Ds \Rt}}"yS";"uS"; 
	\ar^{\, \cmpl{ \mbox{{\sm \dashbox{\dsz}(18, 16)[]{$\SlT' $}}} }  }@/_-1pc/"yS";"xS";
	\ar_{{\Ds \Rr}} @/^-1pc/"yS";"xS"; 
	\endxy}^{{\Ds \SlS'}}  
& \hs{50pt} &
	\underbrace{\xy 
	(50,0)*+{ \inp \, \x' \hs{10 pt}}="xT"; (80,0)*+{ \, \ndv'}="vT"; 
	(60,-15)*+{ \, \ndu'}="uT"; 
	(60,-30)*+{ \hs{10 pt}  \y'  \, \inp}="yT"; 
	\ar_{{\Ds \Rs}}"uT";"xT";  \ar_{{\Ds \Rs}}"uT";"vT"; 
	\ar_{{\Ds \Rt}}"yT";"uT";   \ar^{{\Ds \Rr}}"yT";"vT";
	\endxy}_{{\Ds \SlT'}}  
 \Ea\] 
\item Now consider 	the mapping $\hth$, 
given by 
 $\x', \ndv' \mapsto \x$; $\ndu' \mapsto \z$;  $\y' \mapsto \y$. 
 It maps arcs of $\SlT'$ to arcs of $\SlS'$. 
 Slice $\snt{S}'$ is 
inconsistent, corresponding to the empty relation $\ES$ (see~\ref{subsec:Cnstr}). 
Informally speaking, in $\snt{S}'$ we find an image of $\snt{T}'$ in parallel with $\cmpl{\snt{T}'}$. 
Thus, we have the inclusion 
$\snt{S}' \, \sqeq \, \btm$, whence also the inclusions $\snt{S} \, \sqeq \, \btm$, 
$\RP \, \sa \, \cmpl{\RQ} \, \sqeq \, \btm$ and  $\RP \sqeq \RQ$.  
\Ee 

\end{Exmpl} 

	We will be able to convert every relational term to a graph 
(see~\ref{subsec:Sntx} and~\ref{subsec:Redrls}). 
Consider, however, the following two slices $\SlS'$ and $\SlS''$:  
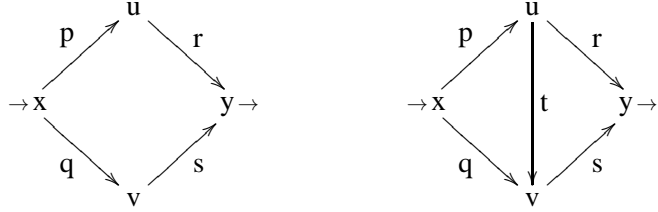
\[ \Ba{ccc}  
	\xymatrix{  & \ndu  \ar[dr]^{{\Ds \Rr}} & \\
	\inp \, \x \ar[ur]^{{\Ds \Rp}} \ar[dr]_{{\Ds \Rq}} &  &  \y  \, \inp \\ 
	 & \ndv  \ar[ur]_{{\Ds \Rs}} &  }  
& \hs{30pt} &
	\xymatrix{  & \ndu  \ar[dr]^{{\Ds \Rr}}  \ar[dd]^{{\Ds \Rt}} & \\
	\inp \, \x \ar[ur]^{{\Ds \Rp}} \ar[dr]_{{\Ds \Rq}} &  &  \y  \, \inp \\ 
	 & \ndv  \ar[ur]_{{\Ds \Rs}} &  }  
 \Ea\] 
Slice $\SlS'$ corresponds to the term 
 $\Lm [ \Ba{c} (\Rp \rp \Rr) \\  \sa \\  (\Rq \rp \Rs) \Ea  \Rm ]$\!, but  
one does not have a term corresponding to slice $\SlS''$. 
So, graphs will turn out to be more expressive than  relational terms. 


\section{Graph Language} \label{sec:GrLng} 
We now introduce  our  graph language:  
syntax and semantics (in~\ref{subsec:Sntx}) and 
some 
constructions  (in~\ref{subsec:Cnstr}).
Labels, slices and graphs will represent binary relations, 
whereas arcs will represent restrictions. 

	We will consider two fixed denumerably infinite sets:  
 set $\RN$ of \emph{relation names} and 
 set $\IN$ of (individual) \emph{nodes} (in alphabetical order: $\nd{x}, \nd{y}, \nd{z}, \dots$). 

\subsection{Syntax and semantics} \label{subsec:Sntx}  
We now examine  the syntax and semantics of our  graph language. 
	
	We introduce  the syntax of our  graph concepts (by mutual recursion). 	

\Bd 	
\item[{\sm ($\lbl{L}$)}] 
The \emph{labels} are  (freely) generated  
from the relation names, 
slices and graphs (see below),  
by relational operations and constants. 
\item[{\sm ($\snt{a}$)}]  
An \emph{arc} 
over a set $N \subseteq \IN$ 
is a triple $\nd{u} \, \lbl{L} \, \nd{v}$, where  $\nd{u}, \nd{v} \in N$ and $\lbl{L}$ is a label. 
\item[{\sm ($\dg$)}] 
A \emph{sketch} $\dg= \BDr N \drs A \EDr$   
consists of $2$ sets: $N \subseteq \IN$ of nodes and $A$ of arcs over $N$. 
\item[{\sm ($\snt{D}$)}] 
A \emph{draft} $\snt{D}$ 
is a sketch with 
finite sets 
of nodes and 
of arcs. 
\item[{\sm ($\snt{S}$)}]  
A \emph{slice} $\SlS = \BSl N \drs A \dps \x_{\SlS}, \y_{\SlS} \ESl$  consists of  
a draft  $\ud{\snt{S}} = \BDr N \drs A \EDr$ (its \emph{underlying draft})   
together with a pair of distinguished nodes $\nd{x}_{\SlS}, \nd{y}_{\SlS} \in N$ (its \emph{input} and \emph{output} nodes). 
For instance, in Example~\ref{Exmpl:RAaxm}, we have the slice   
$\snt{S}_2 = \BSl \{ \x, \y, \z \} \drs \{ \x \, \cnv{\RP} \, \z , \z \, \cmpl{\RP \rp \RQ} \, \y , \x \, \RQ \, \y \}  \dps 
\x, \y \ESl$. 
\item[{\sm ($\snt{G}$)}]  
A graph 
is a finite set of  slices. 
Example~\ref{Exmpl:Exprl} (in~\ref{subsec:Exprl}) will show a $2$-slice graph $\snt{G} = \{ \SlS_+ , \SlS_- \}$. 
\Ed 

	The \emph{empty graph} $\{  \hs{3pt} \}$ has no slice. 
Note that every relational term is a label, as are slices and graphs. 
Drafts, slices and graphs are finite objects, 
whereas sketches are useful in some arguments (cf.~\ref{subsec:Exprl}).

	An \emph{inclusion} is a pair of  labels, noted $\lbl{L} \sqeq \lbl{K}$. 
The \emph{difference slice} of a pair of  labels  $\lbl{L}$ and $\lbl{K}$ 
 is the  $2$-arc slice 
$\ds{\lbl{L}}{\lbl{K}} := \BSl \{ \nd{x}, \nd{y} \} \drs  \{ \nd{x} \,  \lbl{L} \, \nd{y}, \nd{x} \,  \cmpl{\lbl{K}} \, \nd{y} \} \dps \nd{x}, \nd{y} \ESl$    
(where $\nd{x}$ and $\nd{y}$ are the first $2$ nodes in $\IN$). 
The  
difference slice $\ds{\lbl{L}}{\lbl{K}}$ has $2$ parallel arcs: 
$\xy
(0,0)*+{ \inp \, \x \, }="x"; (14,0)*+{ \, \y \, \inp}="y";
{\aro{\lbl{L}}@/_-1pc/"x";"y"};{\aru{\cmpl{\lbl{K}}}@/^-1pc/"x";"y"}
\endxy$ 
(cf. Examples~\ref{Exmpl:RAaxm} and~\ref{Exmpl:Mdlw} in Section~\ref{sec:BsIds}). 

	We now examine  the semantics  of our  graph language. 
We use models for  semantics: a model  assigns a binary relation to each relation name. 
A \emph{model} is a structure  $\gM = \BMd M, \BMr r^{\gM} \EMr_{r \in \RN} \EMd$, 
consisting of a  set $M$ and a binary relation  $r^{\gM}$ on $M$, i.e.  
$r^{\gM} \subseteq M^2$, 
for each relation name $r \in \RN$.  
An \emph{$M$-assignment} for a set $N \subseteq \IN$ of nodes 
is a function  
$\g : N \rar M$, assigning an element  $\ndw^{\g} \in M$ to each  
node $\ndw \in N$. 

	We now  introduce the semantics  of our  graph concepts (again by mutual recursion). 	
Consider a given \emph{$M$-model} $\gM = \BMd M, \BMr r^{\gM} \EMr_{r \in \RN} \EMd$. 

\Bd  
\item[{\sm ($\lbl{L}$)}] 
The \emph{relation of label} $\lbl{L}$ is the relation $\rel{\lbl{L}}_{\gM} \subseteq M^2$ 
obtained by extending the relations of the relation names  by means of the concrete versions of the operations.     
More precisely, the relation of a label  is the binary relation on $M$  defined as follows. 
\Bd 
\item[{\sm ($0$)}]
For a relation name $r$: 
$\rel{r}_{\gM} := r^{\gM}$ (as given by model  $\gM$). 
For the constants, we set $\rel{\, \btm}_{\gM} := \emptyset$,  
 $\rel{\, \tp}_{\gM} := M^2$,  $\rel{\id}_{\gM}:= I_M$ 
 and 
 $\rel{\dv}_{\gM} := I_M \,  \ccmpl{}$. 
 For a slice or  a graph, we employ their extensions, namely:     
	$\rel{\snt{S}}_{\gM} :=  \betM{S}$ and $\rel{\snt{G}}_{\gM} :=  \betM{G}$ 
	(as defined  below).  
\item[{\sm ($1$)}] 
For the unary operations $\cmpl{\strl{}{\hs{\svs pt}}}$ and $\cnv{}$, we have 
	Boolean complementation ~$\, \ccmpl$   and 
	Peircean transposition $\, \trnsp{}$, respectively; so we set 
	$\rel{\cmpl{\lbl{L}}}_{\gM} := {\rel{\lbl{L}}_{\gM}}  \,  \ccmpl{}$ and 
	$\rel{\cnv{\lbl{L}}}_{\gM} := \trnsp{\rel{\lbl{L}}_{\gM}}$. 
\item[{\sm ($2$)}] 
For the binary operations  $\sa$, $\su$,  $\rp$ and  $\rs$, we have 
	intersection,  union, relative product and relative sum, respectively; so we set 
	$\rel{\lbl{L} \sa \lbl{K}}_{\gM} := \rel{\lbl{L}}_{\gM} \cap \rel{\lbl{K}}_{\gM}$, 
	$\rel{\lbl{L} \su \lbl{K}}_{\gM} := \rel{\lbl{L}}_{\gM} \cup \rel{\lbl{K}}_{\gM}$, 
	$\rel{\lbl{L} \rp \lbl{K}}_{\gM} := \rel{\lbl{L}}_{\gM} \mid \rel{\lbl{K}}_{\gM}$ and 
	$\rel{\lbl{L} \rs \lbl{K}}_{\gM} := \rel{\lbl{L}}_{\gM}  \, \ul{{\Ss |}} \, \rel{\lbl{K}}_{\gM}$. 	
\Ed  
\item[{\sm ($\snt{a}$)}]   
An $M$-assignment $\g : N \rar M$ \emph{satisfies} an arc $\nd{u} \, \lbl{L} \, \nd{v}$   in $\gM$ (noted $\g \arcsat{\gM} \nd{u} \, \lbl{L} \, \nd{v}$)  iff 
the pair of values  $\nd{u}^{\g}$ and $\nd{v}^{\g}$ belongs to the relation of the label, i.e.  
$\nd{u}, \nd{v} \in N$ and $\Bop \nd{u}^{\g}, \nd{v}^{\g} \Eop \in \rel{\lbl{L}}_{\gM}$. 
\item[{\sm ($\dg$)}] An assignment $\g : N \rar M$   \emph{satisfies} a sketch  
$\dg = \BDr N_{\dg} \drs A_{\dg} \EDr$   in $\gM$ 
(noted $\g : \dg \rar \gM$) iff it satisfies all its arcs, i.e.   
 $\g \arcsat{\gM} \snt{a}$, for every arc $\snt{a} \in A_{\dg}$.  
\item[{\sm ($\snt{S}$)}] 
	The \emph{extension} of a  slice 
$\SlS = \BSl \ud{\SlS} \dps \x_{\SlS} , \y_{\SlS} \ESl$ is the binary 
relation on $M$ consisting of the pair of values of 
$\nd{x}_{\SlS}$ and $\nd{y}_{\SlS}$ for the  assignments satisfying its underlying draft $\ud{\SlS}$, namely:  
	$$\betM{\snt{S}} := 
	\Setof{\Bop {\nd{x}_{\SlS}}^\g, {\nd{y}_{\SlS}}^\g \Eop \in M^2}{\g : \ud{\SlS} \rar \gM}.$$ 
\item[{\sm ($\snt{G}$)}] 	
The \emph{extension} of a  graph $\snt{G}$  is the union of the extensions of its slices:    
$\betM{\snt{G}} : = \bigcup_{\snt{S} \in \snt{G}} \, \betM{\snt{S}}$.	
\Ed 

\begin{Rem}    \label{Rem:SlDrfsat} 
A slice $\SlS$ has non-empty extension in an $M$-model $\gM$ iff 
some  $M$-assignment satisfies (in $\gM$) its underlying  draft $\ud{\SlS}$. 
\end{Rem}  

	An  inclusion $\lbl{L} \sqeq \lbl{K}$  \emph{holds} in  model $\gM$ 
(noted $\gM \models \lbl{L} \sqeq \lbl{K}$)  
iff 
 $\rel{\lbl{L}}_{\gM} \subseteq \rel{\lbl{K}}_{\gM}$. 
 An inclusion is  \emph{valid} iff it holds in every model. 
 For instance, 
 the inclusions $\cnv{\RP} \rp \, \cmpl{\RP \rp \RQ} \, \sqeq \,  \cmpl{\RQ}$, 
 $( \cnv{\RP} \rp \, \cmpl{\RP \rp \RQ} )  \sa  \cmpl{\cmpl{\RQ}}  \sqeq \, \btm$ and 
 $\snt{S}_i \, \sqeq \, \btm$ (for $i = 0, 1, \dots, 4$)   
 in Example~\ref{Exmpl:RAaxm} 
 are all valid. 
Label $\lbl{L}$ is \emph{null} iff it the inclusion   $\lbl{L} \sqeq \btm$ is valid. 
Clearly, the empty graph $\{  \hs{3pt} \}$ (with no slice) and the constant $\, \btm$ are null. 
 Labels $\lbl{L}$ and $\lbl{K}$ are \emph{equivalent} (noted  $\lbl{L} \eq \lbl{K}$) iff both inclusions 
 $\lbl{L} \sqeq \lbl{K}$ and $\lbl{K} \sqeq \lbl{L}$ are valid. 
 For instance, 
 in Example~\ref{Exmpl:RAaxm}, all slices $\SlS_0$ through $\SlS_4$ are equivalent labels. 
A slice $\snt{S}$ and a singleton graph $\{ \snt{S} \}$ are equivalent, so one may identify them. 

\begin{Lem} \label{Lem:IncDS} 
An  inclusion $\lbl{L} \sqeq \lbl{K}$  \emph{holds} in a model 
$\gM$ ($\gM \models \lbl{L} \sqeq \lbl{K}$)  
iff  the difference slice $\ds{\lbl{L}}{\lbl{K}}$ has empty extension in $\gM$ 
($\betM{\ds{\lbl{L}}{\lbl{K}}} = \ES$). 
 \end{Lem} 
  \begin{proof} 
  The difference slice has extension $\betM{\ds{\lbl{L}}{\lbl{K}}} = 
  \rel{\lbl{L}}_{\gM} \setminus \rel{\lbl{K}}_{\gM}$.   
 \end{proof}  
 
\begin{Cor}    \label{Cor:Slsat} 
An  inclusion $\lbl{L} \sqeq \lbl{K}$  holds in an $M$-model 
 iff  no $M$-assignment 
 satisfies the  underlying  draft of the  difference slice $\ds{\lbl{L}}{\lbl{K}}$. 
\end{Cor}  
 \begin{proof} 
  By Remark~\ref{Rem:SlDrfsat} and Lemma~\ref{Lem:IncDS}.   
 \end{proof}  

\subsection{Concepts and constructions} \label{subsec:Cnstr}
We will now examine some concepts and constructions. 

	We use the notation `$\ada$' for adding arcs to a sketch or to a slice. 
Given an arc 
$\ndu \, \lbl{L} \, \ndv$: for a sketch $\dg = \BDr N \drs A \EDr$, 
$\dg \ada \ndu \, \lbl{L} \, \ndv := \BDr N \cup \{  \ndu , \ndv \}  \drs A  \cup \{ \ndu \, \lbl{L} \, \ndv \} \EDr$;   
for a slice $\SlS$, 
$\SlS \ada \ndu \, \lbl{L} \, \ndv := \BSl \ud{\SlS} \ada \ndu \, \lbl{L} \, \ndv \dps \x_{\SlS} , \y_{\SlS} \ESl$.

	We now introduce  morphisms for comparing sketches. 
	  
	Consider sketches  $\dg' = \BDr N' \drs A' \EDr$  and $\dg'' = \BDr N'' \drs A'' \EDr$. 
A node renaming function $\hth: N' \rar N''$ is a 
\emph{morphism} from $\dg'$ to $\dg''$   
(noted $\hth: \dg' \mor \dg''$) iff it preserves arcs: 
for every arc $\ndu \, \lbl{L}  \, \ndv \in A'$,  $\ndu^{\hth} \, \lbl{L}  \, \ndv^{\hth}$ is an arc in $A''$. 
For instance, Example~\ref{Exmpl:Mdlw} (in Section~\ref{sec:BsIds}) shows a morphism    
$\hth: \ud{\SlT'}  \mor \ud{\SlS'}$. 
We will use $\Mor{\dg'}{\dg''}$ for the \emph{set of morphisms} from $\dg'$ to $\dg''$. 

	Morphisms transfer satisfying assignments by composition. 
	
\begin{Lem} \label{Lem:Hmtrnsfr} 
Given  a   morphism     $\hth: \dg' \mor \dg''$ and a  model  $\gM$, 
for every assignment $\g$ satisfying $\dg''$  in  model  $\gM$,     
the composite  $\g \cmp \hth$ is an assignment satisfying  $\dg'$ in  model  $\gM$.   
\end{Lem}  
\begin{proof}   
  For every arc $\ndu \, \lbl{L}  \, \ndv \in A_{\dg'}$,  we have 
  $\ndu^{\hth} \, \lbl{L}  \, \ndv^{\hth} \in A_{\dg''}$, so   
 $\Bop \ndu^{\g \cmp \hth}, \ndv^{\g \cmp \hth}  \Eop \in \rel{\lbl{L}}_{\gM}$.    
\end{proof}  

 	A sketch $\dg$ is \emph{ zero} iff, for some slice 
$\snt{T} = \BSl \ud{\snt{T}} \dps \x_{\snt{T}} , \y_{\snt{T}} \ESl$, 
there exists a morphism $\hth:  \ud{\snt{T}} \mor \dg$, such that 
${\x_{\snt{T}}}^{\hth} \, \cmpl{\snt{T}}\, {\y_{\snt{T}}}^{\hth}$ is an arc of $\dg$.  
A slice   is \emph{zero} iff its underlying draft  is zero. 
For instance, 
in Example~\ref{Exmpl:Mdlw}, draft $\ud{\SlS'}$  is a zero sketch and slice $\SlS'$  is a zero slice. 
A  \emph{zero graph} is a graph  consisting of zero slices. 


\begin{Lem} \label{Lem:Ansktch} 
No assignment can satisfy a zero sketch.   
\end{Lem} 
 \begin{proof}  
 By  Lemma~\ref{Lem:Hmtrnsfr}.  
If $\g :  \dg \rar \gM$, then we have  $\g \cmp \hth :  \ul{\snt{T}} \rar \gM$ 
 (thus $\Bop {\x_{\snt{T}}}^{\g \cmp \hth}, {\y_{\snt{T}}}^{\g \cmp \hth} \Eop \in \betM{\snt{T}}$)
 and 
 $\g \arcsat{\gM} {\x_{\snt{T}}}^{\hth} \, \cmpl{\snt{T}} \,  {\y_{\snt{T}}}^{ \hth}$ (whence   
 $\Bop {\x_{\snt{T}}}^{\g \cmp \hth}, {\y_{\snt{T}}}^{\g \cmp \hth} \Eop \not \in \rel{\snt{T}}_{\gM}$),  
 giving   a contradiction. 
 \end{proof}  

	Zero graphs  have empty extensions  in every model, thus being null.  
	 
\begin{Cor} \label{Cor:Zrgr} 
A  zero graph $\snt{H}$ 
is null:  $\betM{\snt{H}}  = \emptyset$, for every model $\gM$.  
\end{Cor} 
 \begin{proof} 
 By Remark~\ref{Cor:Slsat}  (in~\ref{subsec:Sntx}) and  Lemma~\ref{Lem:Ansktch}. 
 If   $\betM{\snt{H}} \neq \ES$, then $\betM{\SlT} \neq \ES$, for some slice $\SlT \in \snt{H}$, whence 
 some $M$-assignment satisfies the underlying draft $\ud{\SlT}$.
 \end{proof} 
 
 	We call a model $\gM =  \BMd  M, \BMr r^{\gM} \EMr_{r \in \RN} \EMd$ \emph{natural} for a 
 sketch  $\dg = \BDr N_\dg \drs A_\dg  \EDr$ iff  $M = N$ and, for each   $r \in \RN$, 
$r^{\gM} =  \Setof{ \Bop \ndw, \z \Eop \in {M}^2}{ \ndw \, r \, \z \in A }$. 
For instance, a 
natural model $\gM$ for draft $\ul{\SlS'}$ (in Example~\ref{Exmpl:Mdlw}  in Section~\ref{sec:BsIds}) has 
$M = \{ \x, \y, \z \}$, $\Rr^{\gM} = \{ \Bop \x, \y \Eop \}$, $\Rs^{\gM} = \{ \Bop \x, \z \Eop \}$ and 
$\Rt^{\gM} = \{ \Bop \z, \y \Eop \}$. 
Natural models will be used for establishing completeness (in~\ref{subsec:Exprl}).  

	We will now examine some constructions: co-limits and pushouts~\cite{Mac_71}. 

	We wish to glue a slice $\snt{T}$ onto  a slice  $\snt{S}$ via a designated pair of nodes. 
One can do this as follows. 
\Be 
\item  First,  use identity arcs to connect the input and output nodes of  $\snt{T}$ to  the  designated nodes. One then obtains a slice with the  following aspect: 
 \[ \begin{array}{lcr} 
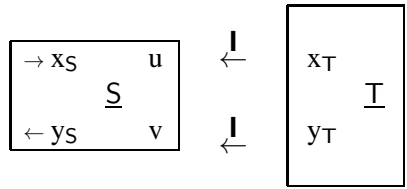
	\begin{array}{|ccc|} 
	\hline  
	\inp \; \x_{\snt{S}} & & \ndu \,  	\\  & \ud{ \snt{S}} & \\ \out \; \y_{\snt{S}} & & \ndv \,  	\\ 
	\hline 
	 \end{array} & 
	 \begin{array}{c} 
	 \stackrel{{\textstyle \id}}{\lar}  \\ 	 \\ \stackrel{{\textstyle \id}}{\lar}  
	 \end{array} 	& 
	 \begin{array}{|cc|} 
	\hline 
	 &  \\\; \x_{\snt{T}} & \\ & \ud{ \snt{T} } \;   \\ \; \y_{\snt{T}} &  \\  & \\ 
	\hline 
	 \end{array} 
 \end{array}  \]  
\item Now,  eliminate the  identity arcs 
to obtain the glued slice 
$\gl{\SlS} {\ndu} {\ndv}  {\SlT}$. 
\Ee 

	We now illustrate this construction. 

	One can eliminate an arc $\ndw \, \id \, \z$ from a slice $\SlS$ by renaming 
	$\ndw$ to $\z$ (or $\z$ to $\ndw$) 
throughout  in $\SlS$. 
For instance, from the 
slice 
$\SlS := \BSl \{ \x, \ndu,  \ndv, \y \}  \drs 
	\{ \x \, \Rt \,  \y ,  \x \, \Rr \,  \ndu ,  \ndu \, \id \,  \ndv ,  \ndv \, \Rs \,  \y \}  \dps \x, \y \ESl$, 
we obtain the (equivalent)  slice 
$\SlS' := \BSl \{ \x,  \ndv, \y \}  \drs 
	\{ \x \, \Rt \,  \y ,  \x \, \Rr \,  \ndv ,  \ndv \, \Rs \,  \y \}  \dps \x, \y \ESl$. 

\begin{Exmpl}  \label{Exmpl:GlSl} 
Consider the three slices $\snt{S} := \Ba{ccccccc} 
	 \inp \, \x  & \strl{{\Ds \Rr}}{\rar} & \ndu  & 
	  \strl{{\Ds \Rs}}{\rar} & \ndv  &  \strl{{\Ds \Rt}}{\rar} &  \y  \,  \inp 
	\Ea$, 
$\snt{T} := \Ba{rcl} \inp  \ndw & \strl{{\Ds \Rp}}{\rar}  & \z \inp \Ea$ and 
$ \xy 
(-10,-0)*+{ \snt{T}' \, := \, \hs{\lvs pt} }="T";
(0,5)*+{ \strl{\dinp \upout}{ \ndw} }="x";
(0,-5)*+{ \z }="z";
{\ar^{{\Ds \Rq}}@/_-1pc/"z";"x"};
{\ar^{{\Ds \Rp}}@/^+1pc/"x";"z"}
\endxy$.  
We then have the following three glued slices:  
$
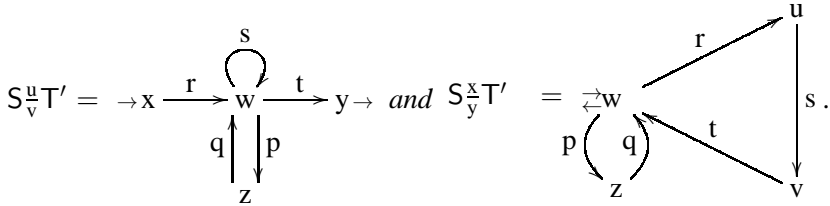
\xy 
(-10,0)*+{  \gl{\SlS} {\ndu} {\ndv}  {\SlT} \, = \hs{\svs pt} }="T";
(0,0)*+{  \inp \, \x}="x";
(10,0)*+{  \ndu}="u"; 
(20,0)*+{  \ndv}="v";
(30,0)*+{  \y \, \inp}="y";
(20,0)*+{}="yS1";
{\ar^{{\Ds \Rr}}@/_-0pc/"x";"u"}; 
{\ar^{{\Ds \Rs}}@/_-1pc/"u";"v"};
{\ar_{{\Ds \Rp}}@/^-1pc/"u";"v"}; 
{\ar^{{\Ds \Rt}}@/_-0pc/"v";"y"}
\endxy $, 
$\gl{\SlS} {\ndu} {\ndv}  {\SlT'}  \, = \, \xymatrix{   \inp \, \x \ar[r]^{{\Ds \Rr}} & 
 \ndw \ar@(ul,ur)[]^{{\Ds \Rs}} \ar@<1ex>[d]^{{\Ds \Rp}} \ar[r]^{{\Ds \Rt}} & 
\y \, \inp \\ 
 & \z \ar@<1ex>[u]^{{\Ds \Rq}} & }$ and 
$ \xy 
(-13,-0)*+{ \gl{\SlS} {\x} {\y}  {\SlT'}   \hs{\lvs pt} =  \hs{\lvs pt} }="T"; 
(-3,0)*+{ \strl{\inp}{\out } \, }="IO";
(0,0)*+{ \, \ndw \hs{5 pt} }="x";
(24,12)*+{ \ndu }="u";
(0,-12)*+{ \z }="z";
(24,-12)*+{\ndv}="v";
{\ar^{{\Ds \Rr}}@/_-0pc/"x";"u"}; 
{\ar^{{\Ds \Rs}}@/_-0pc/"u";"v"}; 
{\ar_{{\Ds \Rt}}@/_-0pc/"v";"x"}; 
{\ar_{{\Ds \Rp}}@/^-1pc/"x";"z"}; 
{\ar^{{\Ds \Rq \, }}@/^-1pc/"z";"x"} 
\endxy$.
\end{Exmpl}  

	The category of sketches and morphisms has co-limits. 
The  co-limit of a diagram of  sketches 
can be obtained as expected: obtain the co-limit of the sets of nodes and then 
transfer arcs (by using the functions to the co-limit node set). 
Thus,  the pushout of drafts gives a draft. 

	Gluing involves an amalgamated sum (of drafts). 
Consider a slice $\SlT$. Given a draft $\snt{D} = \BDr N \drs A \EDr$ and nodes 
$\Bop \ndu, \ndv \Eop \in \IN^2$, 
the \emph{glued draft} $\gl{\snt{D}} {\ndu} {\ndv}  {\SlT}$ is the pushout of drafts $\snt{D}$ and 
$\ud{\SlT}$ over the arcless draft $\BDr \{ \x, \y\}  \drs \ES \EDr$ 
and natural morphisms ($\ha: \x \mpt \ndu , \y \mpt \ndv$ and 
$\hb: \x \mpt \x_{\SlT} , \y \mpt \y_{\SlT}$)  as follows:  
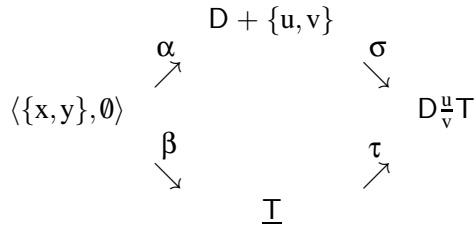
\[ \Ba{ccccc} 
 &  &  \snt{D} \, \ada \, \{ \ndu , \ndv \} &  &  \\ 
  & \strl{{\Ds \ha \, }}{\nearrow} &  & \strl{{\Ds  \, \hsg }}{\searrow} &  \\ 
\BDr \{ \x, \y \}  \drs \ES \EDr &  &  &  &  \gl{\snt{D}} {\ndu} {\ndv}  {\SlT}  \\  
   & \strl{{\Ds \,  \hb }}{\searrow} &  & \strl{{\Ds \hta \, }}{\nearrow} &  \\ 
  &  &  \ud{\snt{T}} &  &
\Ea \]

	Given a slice $\SlS  = \BSl  \ud{\snt{S}} \dps \x_{\snt{S}},  \y_{\snt{S}} \ESl$, we obtain the  
\emph{glued slice} $\gl{\SlS} {\ndu} {\ndv}  {\SlT}$ by transferring the input and output nodes of $\SlS$ to the glued draft $\gl{\ud{\SlS}} {\ndu} {\ndv}  {\SlT}$: 
$\gl{\SlS} {\ndu} {\ndv}  {\SlT} := \BSl  \gl{\ud{\SlS}} {\ndu} {\ndv}  {\SlT} \dps 
	{\x_{\snt{S}}}^\hsg,  {\y_{\snt{S}}}^\hsg \ESl$. 
The glued  draft and slice are unique up to isomorphism.\footnote{As 
isomorphic objects have the same behavior, 
we often consider a sketch or a slice up to isomorphism.} 
Also, we  \emph{glue} a graph naturally by gluing its slices: 
$\gl{\SlS} {\ndu} {\ndv}  {\snt{H}} :=  \Setof{ \gl{\SlS} {\ndu} {\ndv}  {\SlT} }{ \SlT \in \snt{H} }$. 
Note that, for the empty graph: $\gl{\SlS} {\ndu} {\ndv}  {\{  \hs{3pt} \}} =  
\Setof{ \gl{\SlS} {\ndu} {\ndv}  {\SlT} }{ \SlT \in \{  \hs{3pt} \} } = 
\{  \hs{3pt} \}$.  


\section{Refutation Calculus} \label{sec:GrClc} 
We now introduce  our refutation calculus: label conversion and graph expansion. 
We will first examine  basic objects,  
then  rules of  our calculus:  
conversion and its rules (in~\ref{subsec:Redrls}) and 
the expansion rule (in~\ref{subsec:Exprl}). 

	To establish that a label is null, we 
first convert it to a 
graph (by conversion rules) and 
then try to obtain a zero graph by repeatedly applying the expansion rule 
(cf. the examples in Section~\ref{sec:BsIds} and Examples~\ref{Exmpl:Lynd} and~\ref{Exmpl:Exprl}). 

	We define basic labels, arcs, sketches, slices and graphs by mutual recursion.  
A label  $\lbl{L}$ is a \emph{basic label} iff  
it is either a relation name in $\snt{\RN}$ 
or $\cmpl{\snt{T}}$, where $\snt{T}$  is a basic slice (see below).
An arc  $u \, \lbl{L} \, v$ is a \emph{basic arc} iff its  label $\lbl{L}$ is basic. 
A sketch  is a \emph{basic sketch} iff all its arcs  are basic arcs.
A slice  $\SlS$  is a \emph{basic slice} iff
its underlying draft  $\ud{\SlS}$ is a basic sketch. 
A graph    is a \emph{basic graph} iff all its slices   are basic slices.

	In Example~\ref{Exmpl:Mdlw}  (in Section~\ref{sec:BsIds}), 
slice $\snt{S}$ 
is not basic  
(as 
it has composite terms as labels), whereas 
	slice $\snt{S}'$ is basic (as it has $4$ basic arc labels   : $\Rr$, $\Rs$, $\Rt$  and 
$\cmpl{ \SlT' }$, where $\SlT'$ is 
a basic slice).  
Also,  in   Example~\ref{Exmpl:Lynd} (in~\ref{subsec:Redrls} below),   
both slices $\snt{S}$ and $\snt{T}$ are basic.

\subsection{Label conversion} \label{subsec:Redrls}
We now examine label conversion and its rules in our  calculus. 

 \begin{Exmpl}  \label{Exmpl:Lynd} 
Consider the inclusion $\RP \, \sqeq \, \RQ$    (cf.~\cite{Mad_91}), with terms 
$\RP :=  \Ra \,  \sqcap \,   
 	 [  ( \Rb \, \rp \,  \Rc )  \sqcap  \Rd )  
 	  \, \rp \, 
	 (  \Rle  \sqcap ( \Rf \, \rp \,  \Rg) ) ]$  
and 
$\RQ :=    \Rb \,  \rp \, [ ( ( ( \cnv{\Rb} \, \rp \, \Ra )  \sa ( \Rc \, \rp \, \Rle ) ) \rp \, \cnv{\Rg} ) \sa 
				( {\syn{\Rc}{\Rf}} )  \sa 
				( \cnv{\Rb} \, \rp ( ( \Ra \, \rp \, \cnv{\Rg} ) \sa ( \Rd \, \rp \, \Rf )  )  
				] \,  \rp \, \Rg$, 
over relation  names  $\Ra,\Rb,\Rc,\Rd, \Rle,\Rf, \Rg$.  
Label $\RP$ is equivalent to the graph $\{ \SlS \}$, with the following basic slice $ \SlS $:
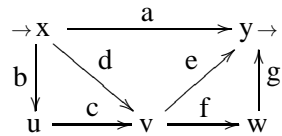
\[  \xymatrix{ 
&
& \inp  \, \x \, \ar[rr] ^{{\Ds \Ra}}  \,  \ar[d] _{{\Ds \Rb}} \,  \ar[dr] ^{{\Ds \Rd}} &&  \y  \,   \inp
\\
&& \ndu\, \ar[r] ^{{\Ds \Rc}} &\ndv \, \ar[ur] ^{{\Ds \Rle}} \, \ar[r] ^{{\Ds \Rf}}& \ndw \, \ar[u] _{{\Ds \Rg}}	
} \]
Label $\RQ$ is equivalent to the graph $\{ \SlT \}$, with the following basic slice $\SlT$:  		
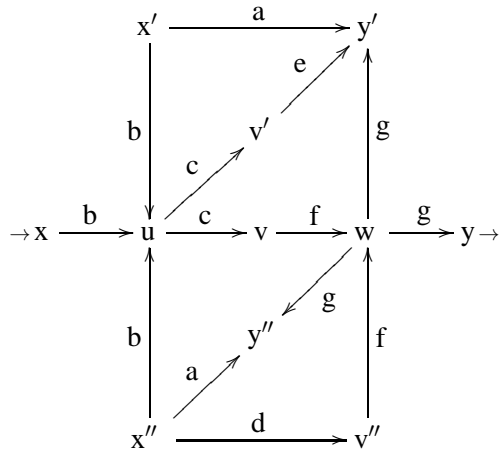
	\[  \xymatrix{ 
& & &  \x' \, \ar[dd] _{{\Ds \Rb}} \ar[rr] ^{{\Ds \Ra }}  && \y' & & 
\\
& & & & \ndv'\, \ar[ur]^{{\Ds \Rle}}
\\
 & & \inp \, \x  \, \ar[r]^{{\Ds \Rb}} & \ndu \, \ar[r]^{{\Ds \Rc}}  \ar[ur]^{{\Ds \Rc}} & \ndv  \ar[r]^{{\Ds \Rf}}& 
 \ndw \, \ar[r]^{{\Ds \Rg}}  \ar[dl]^{{\Ds \Rg}}  \, \ar[uu]_{{\Ds \Rg}}& \y  \, \inp
 \\
 & & & & \y'' &  &  
 \\
 &&&  \x'' \, \ar[rr]^{{\Ds \Rd}}  \, \ar[ur]^{{\Ds \Ra}} \, \ar[uu]^{{\Ds \Rb}} && \ndv'' \ar[uu]_{{\Ds \Rf}} &
}  \]
 So, the difference slice   $\ds{\RP}{ \RQ }$  is equivalent to the graph 
$\{ \SlS \,  \ada \, \x \, \cmpl{\SlT} \, \y \}$. 
Now, we have a morphism $\hth : \ud{\SlT} \mor \ud{\SlS}$ 
given by $\x, \x', \x'' \mpt \x$; $\ndu \mpt \ndu$; $\ndv, \ndv', \ndv'' \mpt \ndv$;  
$\ndw \mpt \ndw$ and $\y, \y', \y'' \mpt \y$. 
Thus, $\{ \SlS \,  \ada \, \x \, \cmpl{\SlT} \, \y \}$ is a zero graph, so inclusions 
$\{ \SlS \,  \ada \, \x \, \cmpl{\SlT} \, \y \} \, \sqeq \, \btm$, 
$\ds{\RP}{ \RQ } \, \sqeq \, \btm$ and $\RP \, \sqeq \, \RQ$ are all valid.  

\end{Exmpl}   

	The conversion rules will be  of two kinds: operational  and structural rules.  
The aim of these rules is converting every label to an equivalent basic graph (see Proposition~\ref{Prop:Red}). 
 
	The operational rules come from labels  that are  equivalent to  graphs.\footnote{Recall 
that $\x$, $\y$ and $\z$ are the first $3$ individual nodes (see Section~\ref{sec:GrLng}).}  
For the constants:  
$\btm$ is equivalent to the empty graph   $\{  \hs{3pt} \}$, 
$\tp$ and  $\id$ are equivalent to graphs with a single arcless slice, namely 
$\inp \, \x   \hs{5pt}  \y \, \inp$ and $\inp \, \x \, \inp$;  
also, for diversity $\dv \eq 
\{
\xy
(0,0)*++{\ \inp \, \nd{x}}="xS";
(20,0)*++{ \nd{y} \, \inp}="yS2";
{\ar^{ \cmpl{{\Ds \mbox{ \dashbox{\dsz}(28, 18)[]{$\inp \, \x \,  \inp$}}} }  }"xS";"yS2"}
\endxy
\}$.  
For the operations:  
$\cnv{\lbl{L}} \eq  \{  \inp \, \x \, \strl{{\Ds \lbl{L} }}{\lar} \,  \y \, \inp \}$, 
$\lbl{L} \sa \lbl{K}$ is   equivalent to the graph whose single slice consists of  the $2$ parallel arcs 
$\inp \, \x \, \strl{{\Ds \lbl{L} }}{\rar} \,  \y \, \inp$ and $\inp \, \x \, \strl{{\Ds \lbl{K} }}{\rar} \,  \y \, \inp$,  
$\lbl{L} \su \lbl{K}$ is   equivalent to the 
graph $\{  \inp \, \x \, \strl{{\Ds \lbl{L} }}{\rar} \,  \y \, \inp , \inp \, \x \, \strl{{\Ds \lbl{K} }}{\rar} \,  \y \, \inp \}$, 
$\lbl{L} \rp \lbl{K}$ is   equivalent to the graph with single slice 
$\inp \, \x \, \strl{{\Ds \lbl{L}}}{\rar} \,  \z   \, \strl{{\Ds \lbl{K}}}{\rar} \,  \y \, \inp$ and 
$\lbl{L} \rs \lbl{K}$ is   equivalent to the graph with single slice  
$\xy 
(0,0)*+{ \inp \, \x \hs{\svs pt}}="x"; 
(44,0)*+{ \hs{\svs pt}  \nd{y}  \, \inp}="y"; 
\ar^{{\Ds \cmpl{ \mbox{ \dashbox{\dsz}(80, 22)[]{$\inp \, \x \, \strl{{\Ds \cmpl{\lbl{L}}}}{\rar} \,  \z   \, \strl{{\Ds \cmpl{\lbl{K}}}}{\rar} \,  \y \, \inp$}}} \, }}"y";"x";
\endxy $.  
We have no such rule for complementation, but we do have $\cmpl{\cmpl{\lbl{L}}}  \eq    \lbl{L}$. 
We will consider  the \emph{consecutive-arc slice} $\CS{\lbl{L}}{\lbl{K}} := \BSl \{ \x, \y, \z \} \drs \{ \x \, \lbl{L} \, \z , \z \, \lbl{K} \, \y \} \dps \x , \y \ESl$,  
i.e.  the slice $\inp \, \x \, \strl{{\Ds \lbl{L}}}{\rar} \,  \z   \, \strl{{\Ds \lbl{K}}}{\rar} \,  \y \, \inp$.

	Table~\ref{Tab:OpLblRls} gives the $10$ \emph{operational rules}. 
\begin{table}[htbp] 
\[ \Ba{lcccl}
\mbox{{\sm ($\btm$)}} & \btm & \red &\{  \hs{3pt} \} & 
	\mbox{{\small empty graph}}  \vs{\nvs pt} \\ 
\mbox{{\sm ($\tp$)}} & \tp & \red & \{ \BSl \{ \x, \y \} \drs \ES \dps \x , \y \ESl \} & 
	\mbox{{\small $2$-node arcless slice $\inp \, \x   \hs{5pt}  \y \, \inp$}}  \vs{\mvs pt} \\ 
\mbox{{\sm ($\id$)}} & \id & \red & \{ \BSl \{ \x \} \drs \ES \dps \x , \x \ESl \} & 
	\mbox{{\small single-node arcless slice $\inp \, \x \, \inp$}}  \vs{\nvs pt} \\ 
\mbox{{\sm ($\dv$)}} & \dv & \red &  
	\{  \BSl \{ \x, \y \} \drs \{ \cmpl{\BSl \{ \x \} \drs \ES \dps \x , \x \ESl} \}  \dps \x , \y \ESl \} & 
	\mbox{{\small $2$-node single-arc slice 
	${\Ss \xy
(0,0)*++{\ \inp \, \nd{x}}="xS";
(20,0)*++{ \nd{y} \, \inp}="yS2";
{\ar^{ \cmpl{{\Ds \mbox{ \dashbox{\dsz}(28, 18)[]{$\inp \, \x \,  \inp$}}} }  }"xS";"yS2"}
\endxy }$}}  \vs{\lvs pt} \\ 
 \mbox{{\sm ($\dc$)}} &  \cmpl{\cmpl{\lbl{L}}} & \red &   \lbl{L} & 
	\mbox{{\small replace $\cmpl{\cmpl{\lbl{L}}}$  by $\lbl{L}$}}  \vs{\mvs pt} \\ 
\mbox{{\sm ($\cnv{}$)}} & \cnv{\lbl{L}} & \red &  
	\{ \BSl \{ \x, \y \} \drs \{ \y \, \lbl{L} \, \x \} \dps \x , \y \ESl \} & 
	\mbox{{\small reversed-arc slice $\inp \, \x \, \strl{{\Ds \lbl{L} }}{\lar} \,  \y \, \inp$}}  \vs{\lvs pt} \\ 
\mbox{{\sm ($\sa$)}} & \lbl{L} \sa \lbl{K}  & \red &   
	\{ \BSl \{ \x, \y \} \drs \{ \x \, \lbl{L} \, \y , \x \, \lbl{K} \, \y \} \dps \x , \y \ESl \} & 
	\mbox{{\small parallel-arc slice:  
	${\Ss \xy
(0,0)*++{ \inp \, \x \, }="x";
(24,0)*++{\, \y \, \inp}="y";
{\ar^{{\Ds \lbl{L}}}@/_-1pc/"x";"y"};
{\ar_{{\Ds \lbl{K}}}@/^-1pc/"x";"y"}
\endxy }
$}}  \vs{\nvs pt} \\ 
\mbox{{\sm ($\su$)}} & \lbl{L} \su \lbl{K}  & \red &  
	\Lm \{ \Ba{c} \BSl \{ \x, \y \} \drs \{ \x \, \lbl{L} \, \y  \} \dps \x , \y \ESl \, , \\  
	\BSl \{ \x, \y \} \drs \{ \x \, \lbl{K} \, \y \} \dps \x , \y \ESl  \Ea \Rm  \} & 
	\mbox{{\small alternative slices: 
	$\Ba{c} \inp \, \x \, \strl{{\Ds \lbl{L} }}{\rar} \,  \y \, \inp \\  
	\inp \, \x \, \strl{{\Ds \lbl{K} }}{\rar} \,  \y \, \inp \Ea$}}  \vs{\mvs pt} \\  
\mbox{{\sm ($\rp$)}} & \lbl{L} \rp \lbl{K}  & \red & 
	\{ \CS{\lbl{L}}{\lbl{K}} \} & 
	\mbox{{\small consecutive-arc slice:  
	$\inp \, \x \, \strl{{\Ds \lbl{L}}}{\rar} \,  \z   \, \strl{{\Ds \lbl{K}}}{\rar} \,  \y \, \inp$}}  \vs{\nvs pt} \\ 
\mbox{{\sm ($\rs$)}} & \lbl{L} \rs \lbl{K}  & \red &
	 \{ \BSl \{ \x, \y \} \drs \{ \x \, \cmpl{\CS{\cmpl{\lbl{L}}}{\cmpl{\lbl{K}}}} \, \y \} \dps \x , \y \ESl  \} & 
	\mbox{{\small complemented  
	label: 
	$\cmpl{\inp \, \x \, \strl{{\Ds \cmpl{\lbl{L}}}}{\rar} \,  \z   \, \strl{{\Ds \cmpl{\lbl{K}}}}{\rar} \,  \y \, \inp}$}}  
\Ea \]
\caption{Operational rules} \label{Tab:OpLblRls}
\end{table} 

	By applying the  operational rules (of Table~\ref{Tab:OpLblRls})  
	in any context, one can  
 eliminate all relational constants and operations except complement,  
 but   complemented relation names (e.g  $\cmpl{\Rr}$) remain and 
  slices or graphs and their complements as labels may appear.   
 
\begin{Exmpl}  \label{Exmpl:OpRls}  
 The  operational rules (in Table~\ref{Tab:OpLblRls})  give the following conversions. 
\Be 
\item  $ \Rr \, \rp \, \btm \,  \cnvu{\rp}  \, 
 	\{  \inp \, \x \, \strl{{\Ds \Rr}}{\rar} \, \z  \, \strl{{\Ds \btm }}{\rar} \,  \y \, \inp   \} \, \cnvu{\btm} \, 
	\Lm \{ \Ba{c} \inp \, \x \, \strl{{\Ds \Rr}}{\rar} \,  \z \, \strl{{\Ds \mbox{ \dashbox{\dsz}(20, 22)[]{
		$\{   \hs{3pt} \}$}}}}{\lngrar}  \, y \, \inp  \Ea \Rm  \} \,  = \,  \snt{G}_1  $.
\item  $ \Rr \, \rp \,  (\Rs  \su \Rt) \,  \cnvu{\rp}  \, 
 	\{  \inp \, \x \, \strl{{\Ds \Rr}}{\rar} \, \z  \, \strl{{\Ds \Rs  \su \Rt }}{\rar} \, \y \, \inp   \} \, \cnvu{\su}  \, 
\Lm \{ \Ba{c}  \xy
(0,0)*+{ \inp \, \nd{x}}="xS";
(10,0)*+{  \nd{z}}="zS2";
{\ar^{{\Ds \Rr}}"xS";"zS2"};
(48,0)*++{ \nd{y} \, \inp}="yS2";
{\ar^ (.4){
\mbox{ \dashbox{\dsz}(76, 40)[]{
		$\Lm \{ \Ba{c}  \inp \, \x \, \strld{ \Rs}{\rar} \, \y \, \inp\, ,  \\ \inp \, \x \, \strld{ \Rt }{\rar} \, \y \, \inp 
		\Ea \Rm  \}$}} 
}"zS2";"yS2"}
\endxy \Ea \Rm  \} \,  = \,  \snt{G}_2$. 
\item  $ \Rr \, \rp \,  \cmpl{\Rs  \sa \Rt} \,  \cnvu{\rp}  \, 
 	\{  \inp \, \x \, \strl{{\Ds \Rr}}{\rar} \, \z  \, \strl{{\Ds \cmpl{\Rs  \sa \Rt} }}{\rar} \, \y \, \inp  \} 
	\, \cnvu{\sa}  \, 
\Lm \{ \Ba{c} \xy
(0,0)*+{ \inp \, \nd{x}}="xS";
(10,0)*+{  \nd{z}}="zS2";
{\ar^{{\Ds \Rr}}"xS";"zS2"};
(48,0)*++{ \nd{y} \, \inp}="yS2";
{\ar^ (.4){
\cmpl{{\Ds \mbox{  \dashbox{\dsz}(68, 42)[]{
$\Lm \{  \Ba{c} 
\xy
(0,0)*+{ \inp \, \nd{x} \, }="xS";
(16,0)*+{ \, \nd{y} \, \inp}="yS2";
{\ar^{{\Ds \Rs}}@/_-1pc/"xS";"yS2"};
{\ar_{{\Ds \Rt}}@/^-1pc/"xS";"yS2"}
\endxy
\Ea \Rm \}$	}}}}
}"zS2";"yS2"}
\endxy  \Ea \Rm  \} 
	\, = \, \snt{G}_3$.
\Ee
\end{Exmpl}  

	The $4$ structural rules ($\strl{ \cup}{\rar}$), ($\cmpl{\cup}$),  ($\cmpl{\cap}$) and  ($\cmpl{r}$) 
will address such cases.\footnote{Recall that a slice $\SlS$ and its 
	single-slice graph  $\{ \SlS \}$ are equivalent (cf.~\ref{subsec:Sntx}).} 
\Bi 
\item[{\sm ($\strl{ \cup}{\rar}$)}] We can replace a graph arc by glued slices (cf.~\ref{subsec:Cnstr}),  as   
$ \SlS \, \ada \,  \ndu \, \snt{H} \, \ndv \,  \eq \, 
 \Setof{  \gl{\SlS} {\ndu} {\ndv}  {\SlT} }{ \SlT \in \snt{H} }$. 
For instance, with the 
slices 
$\SlS := \BSl \{ \x, \ndu, \ndv, \y \} \drs \{ \x \, \Rr \, \ndu , \ndu \, \Rs \, \ndv , \ndv \, \Rt \, \y \}   \dps \x, \y \ESl$, 
$\SlT := \BSl \{ \ndw , \z \} \drs \{ \ndw \, \Rp \, \z \}   \dps \ndw, \z \ESl$ and 
$\SlT' := \BSl \{ \ndw , \z \} \drs \{ \ndw \, \Rp \, \z ,  \z \, \Rq \, \ndw \}   \dps \ndw, \ndw \ESl$ 
(cf. Example~\ref{Exmpl:GlSl} in  \ref{subsec:Cnstr}), we have  
$\SlS \, \ada \,  \ndu \, \{ \SlT , \SlT'\}  \, \ndv$ equivalent to 
$\{ \BSl \{ \x, \ndu, \ndv, \y \} \drs \{ \x \, \Rr \, \ndu , \ndu \, \Rs \, \ndv , \ndu \, \Rp \, \ndv ,  \ndv \, \Rt \, \ndv \}   \dps \x, \y \ESl , 
 \BSl \{ \x, \ndv, \z, \y \} \drs \{ \x \, \Rr \, \ndv , \ndv \, \Rs \, \ndv , \ndv \, \Rt \, \y , \ndv \, \Rp \, \z ,  
 \z \, \Rq \, \ndv \}   \dps \x, \y  \ESl \}$. 
 
\item[{\sm ($\cmpl{\cup}$)}] Also, we can replace a label that is a complemented graph  by a slice,  since   
$\cmpl{\snt{G}} \, \eq \, \SlG{\snt{G}}$, 
where $\SlG{\snt{G}} := \BSl \{ \x, \y \} \drs \Setof{\x \, \cmpl{ \SlS} \, \y}{\SlS \in \snt{G}} \dps \x, \y \ESl$
is the \emph{slice of graph} $\snt{G}$. 
For 
a $2$-slice graph $\snt{G} = \{ \SlS_1 , \SlS_2 \}$, 
$\SlG{\snt{G}}$ is the $2$-arc slice  
$\xy
(0,0)*+{ \inp \, \x \, }="x"; 
(14,0)*+{ \, \y \, \inp}="y";
{\aro{ \cmpl{\SlS_1}}@/_-1pc/"x";"y"};  
{\aru{\cmpl{ \SlS_3 }}@/^-1pc/"x";"y"}
\endxy$. 
\item[{\sm ($\cmpl{\cap}$)}] 
Consider a slice $\SlS = \BSl N \drs A \dps \x_{\SlS}, \y_{\SlS} \ESl$. 
 Call slice $\SlS$ \emph{small} iff $N = \{ \x_{\SlS}, \y_{\SlS} \}$.  
An \emph{I-O arc} of $\SlS$ is an arc $\ndu \, \lbl{L} \, \ndv \in A$ with 
$\{ \ndu ,  \ndv \} \incl\{\x_{\SlS}, \y_{\SlS} \}$. The \emph{transformed} of I-O arc 
$\snt{a} = \ndu \, \lbl{L} \, \ndv$ is the arc $\tra{\snt{a}}$ obtained by replacing $\x_{\SlS}$ by $\x$,  $y_{\SlS}$ by $\y$ and  
label $ \lbl{L}$ by $ \cmpl{\lbl{L}}$. 
Now, the \emph{graph of slice} $\SlS$ is the graph $\GrS{\SlS}$ with  a single-arc slice 
$\BSl \{ \x , \y \} \drs \{ \tra{\snt{a}} \} \dps \x, \y \ESl$, for each I-O arc $\snt{a}$ of $\SlS$. 
For 
a $3$-arc small slice 
$\SlS = \BSl \{ \ndw , \z \} \drs \{ \ndw \, \Rr \, \z ,  \z \, \Rs \, \ndw , \ndw \, \Rt \, \ndw \}   \dps \ndw, \z \ESl$, 
$\GrS{\SlS}$ is a   graph with $3$ slices, namely  
$\BSl \{ \x , \y \} \drs \{ \x \, \cmpl{\Rr} \, \y  \}   \dps \x, \y \ESl$, 
$\BSl \{ \x , \y \} \drs \{ \y \, \cmpl{\Rs} \, \x  \}   \dps \x, \y \ESl$ and 
$\BSl \{ \x , \y \} \drs \{ \x \, \cmpl{\Rt} \, \x  \}   \dps \x, \y \ESl$; 
pictorially, we have 
$\GrS{
	\xymatrix{  
\inp \,  \ndw  \ar@(ul,ur)[]^{{\Ds \Rt}} \ar@<1ex>[r]^{{\Ds \Rr}} & 
\z \, \inp \ar@<1ex>[l]^{{\Ds \Rs}} } }$ as  the graph  
$ \{ \inp \x \, \strld{\cmpl{\Rr}}{\rar} \, \y \inp ,  \inp \x \, \strld{\cmpl{\Rs}}{\lar} \, \y \inp , 
 \inp \,  \strl{{\Ds \strl{{\Ds \cmpl{\Rt}}}{\circlearrowleft}}}{\x}  \, \y \inp \}$. 
Now, for a small slice,  we can replace the complemented slice by a graph,  moving complement inside, 
as $\cmpl{\{ \SlS \}} \eq  \GrS{\SlS}$.
 
\item[{\sm ($\cmpl{r}$)}] Finally, we can replace a label $\cmpl{r}$ by $\cmpl{\inp \, \x \, \strl{{\Ds r}}{\rar}  \,  \y \, \inp}$ 
(since  $\cmpl{\lbl{L}} \, \eq \, \cmpl{\inp \, \x \, \strl{{\Ds \lbl{L}}}{\rar} \,  \y \, \inp}$). 
\Ei 	

\begin{Exmpl}  \label{Exmpl:StrRls} 
The graphs $\snt{G}_1$, $\snt{G}_2$ and $\snt{G}_3 $ in Example~\ref{Exmpl:OpRls}  
have conversions as follows. 
\Be 
\item For $\snt{G}_1$: 
	$\Lm \{ \Ba{c} \inp \, \x \, \strl{{\Ds \Rr}}{\rar} \,  \z \, 
	 \strl{{\Ds \mbox{ \dashbox{\dsz}(20, 22)[]{
	$\{   \hs{3pt} \}$}}}}{\lngrar}  \, \y \, \inp  \Ea \Rm  \} \,  \cnvu{\strl{ \cup}{\rar}}  \, 
	\Setof{  \inp \, \x \, \strl{{\Ds \Rr}}{\rar} \,  \z \, 
	 \strl{{\Ds \mbox{ \dashbox{\dsz}(16, 20)[]{
	$\SlT$}}}}{\lngrar}  \, y \, \inp  }{ \SlT \in \{  \hs{3pt}  \} } \, = \,  \{  \hs{3pt}  \}$. 
\item For $\snt{G}_2$: 
	$\Lm \{ \Ba{c}   \xy
(0,0)*+{ \inp \, \nd{x}}="xS";
(10,0)*+{  \nd{z}}="zS2";
{\ar^{{\Ds \Rr}}"xS";"zS2"};
(48,0)*++{ \nd{y} \, \inp}="yS2";
{\ar^ (.4){
\mbox{ \dashbox{\dsz}(76, 40)[]{
		$\Lm \{ \Ba{c}  \inp \, \x \, \strld{ \Rs}{\rar} \, \y \, \inp\, ,  \\ \inp \, \x \, \strld{ \Rt }{\rar} \, \y \, \inp \Ea \Rm  \}$}} 
}"zS2";"yS2"}
\endxy \Ea \Rm  \}  \,  \cnvu{\strl{ \cup}{\rar}}  \, 
\Lm \{ \Ba{c}  \inp \, \x \, \strl{{\Ds \Rr}}{\rar} \, \z \, \strl{{\Ds \Rs }}{\rar} \, \y \, \inp \, , \\   
	 \inp \, \x \, \strl{{\Ds \Rr}}{\rar} \, \z \, \strl{{\Ds \Rt }}{\rar} \,  \y \, \inp \Ea \Rm  \} $. 
\item For $\snt{G}_3$: 
$\Lm \{ \Ba{c} \xy
(0,0)*+{ \inp \, \nd{x}}="xS";
(10,0)*+{  \nd{z}}="zS2";
{\ar^{{\Ds \Rr}}"xS";"zS2"};
(48,0)*++{ \nd{y} \, \inp}="yS2";
{\ar^ (.4){
\cmpl{{\Ds \mbox{  \dashbox{\dsz}(68, 46)[]{
$\Lm \{  \Ba{c} 
\xy
(0,0)*+{ \inp \, \nd{x} \, }="xS";
(16,0)*+{ \, \nd{y} \, \inp}="yS2";
{\ar^{{\Ds \Rs}}@/_-1pc/"xS";"yS2"};
{\ar_{{\Ds \Rt}}@/^-1pc/"xS";"yS2"}
\endxy
\Ea \Rm \}$	}}}}
}"zS2";"yS2"}
\endxy  \Ea \Rm  \} \,  \cnvu{\cmpl{\cap}} \, 
\Lm \{ \Ba{c} \xy
(0,0)*+{ \inp \, \nd{x}}="xS";
(10,0)*+{  \nd{z}}="zS2";
{\ar^{{\Ds \Rr}}"xS";"zS2"};
(48,0)*++{ \nd{y} \, \inp}="yS2";
{\ar^ (.4){
\mbox{ \dashbox{\dsz}(78, 40)[]{
		$\Lm \{ \Ba{c}  \inp \, \x   \, \strl{{\Ds \cmpl{\Rs} }}{\rar} \, \y \, \inp \, ,  \\   
	 \inp \, \x \,  \strl{{\Ds \cmpl{\Rt} }}{\rar} \, \y \, \inp \Ea \Rm \}$}} 
}"zS2";"yS2"}
\endxy  \Ea \Rm  \} $\\
$  \cnvu{\strl{ \cup}{\rar}} \, 
 \Lm \{ \Ba{c}   \inp \, \x \, \strl{{\Ds \Rr}}{\rar} \, \z \, \strl{{\Ds \cmpl{\Rs} }}{\rar} \, \y \, \inp \,  , \\   
	 \inp \, \x \, \strl{{\Ds \Rr}}{\rar} \, \z \, \strl{{\Ds \cmpl{\Rt} }}{\rar} \, \y \, \inp \Ea \Rm \} 
\, \cnvu{\cmpl{r}} \, 
\Lm \{ \Ba{c} \inp \, \x \, \strl{{\Ds \Rr}}{\rar} \,   \xymatrix@C60pt{ 
	 \z \ar[r]^{{\Ds \cmpl{{\Ds \mbox{ \dashbox{\dsz}(50, 28)[]{
		$ \inp \, \x  \strl{{\Ds \Rs }}{\rar}  \y \, \inp $}}}} }} &  \y \, \inp } \, , \\    
	\inp \, \x \, \strl{{\Ds \Rr}}{\rar} \, \xymatrix@C60pt{ 
	 \z \ar[r]^{{\Ds \cmpl{{\Ds \mbox{ \dashbox{\dsz}(50, 28)[]{
		$ \inp \, \x  \strl{{\Ds \Rt }}{\rar}  \y \, \inp  $}}}} }} &  \y \, \inp } \Ea \Rm \}$. 
\Ee 
\end{Exmpl}  

	Table~\ref{Tab:StrlRls} gives the $4$ \emph{structural rules}. 
\begin{table}[htbp] 
\[ \Ba{lcccl} 
 \mbox{{\sm ($\strl{ \cup}{\rar}$)}} & 
 \{ \, \SlS \, \ada \,  \ndu \, \snt{H} \, \ndv  \, \} & \red & 
 	 \gl{\SlS} {\ndu} {\ndv}  {\snt{H} } & 
	\mbox{{\sm  replace graph arc  by glued slices}}  \vs{\svs pt} \\ 
\mbox{{\sm ($\cmpl{\cup}$)}}   & \cmpl{\snt{G}} & \red & 
	\SlG{\snt{G}}  & 
	\mbox{{\small replace $\cmpl{\snt{G}}$  by slice of $\snt{G}$}}  \vs{\svs pt} \\ 
\mbox{{\sm ($\cmpl{\cap}$)}} \hs{\svs pt}   \mbox{small $\SlS$}  &  \cmpl{\{ \SlS \}} & \red &
	 \GrS{\SlS}   & 
	\mbox{{\sm replace $\cmpl{\{ \SlS \}}$  by graph of $\SlS$}}  \vs{\svs pt} \\ 
 \mbox{{\sm ($\cmpl{r}$)}} \hs{\svs pt}   \mbox{$r \in \RN$} &  \cmpl{r}  & \red & 
 	 \cmpl{\BSl \{ \x, \y \} \drs \{ \x \, r \, \y \} \dps \x , \y \ESl}    & 
	\mbox{{\sm  replace 
	$\cmpl{r}$ by 
	label $\cmpl{\inp \, \x \, \strl{{\Ds  r }}{\rar} \,  \y \, \inp}$}} 
\Ea \]
\caption{Structural rules} \label{Tab:StrlRls}
\end{table} 

	We also have a derived rule replacing 
a complemented graph arc  
by parallel complemented slice arcs: 
\[ \Ba{lcccl} 
 \mbox{{\sm ($\strl{ \cmpl{\cup}}{\rar}$)}} & \{ \, \SlS \, \ada \,  \ndu \, \cmpl{\snt{H}} \, \ndv \, \} & \red & 
 	\{ \, \SlS \, \ada  \,\Setof{  \ndu \, \cmpl{\snt{T}} \, \ndv  }{ \SlT \in \snt{H} }  \, \}  & 
	\mbox{{\small  replace $\ndu \,  \strl{{\Ds \cmpl{\snt{H}}}}{\rar} \, \ndv$ by  
	$\Setof{  \ndu \, \strl{{\Ds \cmpl{\snt{T}}}}{\rar} \, \ndv  }{ \SlT \in \snt{H} }$}} 
\Ea \] 

	Derived rule ($\strl{ \cmpl{\cup}}{\rar}$) is obtained by applying rules $(\cmpl{\cup})$ and 
$(\strl{ \cup}{\rar})$ as follows: 
\[ \Ba{ccc} 
\Lm \{ \Ba{c} \SlS \, \ada \, \xy (0,0)*+{  \nd{u}}="u"; (26, 0)*++{ \nd{v} }="v"; 
{\ar^ (.4){ \hs{10 pt} 
\cmpl{{\Ds  \mbox{  \dashbox{\dsz}(50, 82)[]{$\Lm \{  
\Ba{c} \SlT_1,  \\ {\SSs \vdots}  \\ \SlT_i,  \\  {\SSs \vdots}  \\ \SlT_n \Ea \Rm \}$}}}}
}"u";"v"} 
\endxy \Ea \Rm \}   &  \cnvu{\cmpl{\cup}}  & 
\Lm \{ \Ba{c} \SlS \, \ada \, \xy
(0,0)*+{  \nd{u}}="u"; (37,0)*++{ \nd{v} }="v";
{\ar^ (.4){{\Ds \hs{15 pt} \mbox{  \dashbox{\dsz}(76, 76)[]{$ \xy
(0,0)*+{ \inp \, \nd{x} \, }="xT"; (20,0)*+{ \, \nd{y} \, \inp}="yT"; 
(10,6)*+{{\SSs \vdots}};  (10,-3)*+{{\SSs \vdots}}; 
{\ar^{{\Ds \cmpl{\SlT_1}}}@/_-1.8pc/"xT";"yT"}; {\ar|{{\Ds \cmpl{\SlT_i}}}@/_0pc/"xT";"yT"};
{\ar_{{\Ds \cmpl{\SlT_n}}}@/^-1.8pc/"xT";"yT"}
\endxy $ }}}}"u";"v"}
\endxy  \Ea \Rm \} \vs{\lvs pt} \\ 
& \cnvu{\strl{ \cup}{\rar}}   & 
\Lm \{ \Ba{c} \SlS \, \ada  \,  \ndu \, \strl{{\Ds \cmpl{\snt{T_1}}}}{\rar} \, \ndv   
\,  \ada  \, \dots  \ada  \,  \ndu \, \strl{{\Ds \cmpl{\snt{T_i}}}}{\rar} \, \ndv  
 \ada   \, \dots  \ada  \,  \ndu \, \strl{{\Ds \cmpl{\snt{T_n}}}}{\rar} \, \ndv \,  \Ea \Rm \}
\Ea \]

	The $14$ \emph{conversion rules} (in Tables~\ref{Tab:OpLblRls} and~\ref{Tab:StrlRls}) 
can be applied in any context. 
 We take  the \emph{eventual conversion} relation $\clrd$ as the 
 reflexive-transitive closure of the immediate conversion relation $\red$ 
 under relational operations as well as  slice and graph formation. 
 More precisely:  
 if $\lbl{L} \clrd \lbl{K}$ then $\cmpl{\lbl{L}} \clrd \cmpl{\lbl{K}}$, $\cnv{\lbl{L}} \clrd \cnv{\lbl{K}}$ and 
$\SlS \, \ada \,   \ndu \, \lbl{L} \, \ndv \clrd \SlS \, \ada \,   \ndu \, \lbl{K} \, \ndv$; 
if $\lbl{L}_1 \clrd \lbl{K}_1$ and $\lbl{L}_2 \clrd \lbl{K}_2$ then 
$\lbl{L}_1 \gop \lbl{L}_2 \clrd  \lbl{K}_1 \gop \lbl{K}_2$ 
(for a $2$-ary operation $\gop \in \{ \su , \sa , \rp , \rs \}$); 
if $\SlT \clrd \SlT'$ then $\snt{G} \cup \{ \SlT \} \clrd \snt{G} \cup \{ \SlT' \}$ and 
if $\snt{H} \clrd \snt{H} '$ then $\snt{G} \cup \snt{H}  \clrd \snt{G} \cup \snt{H}'$.

	One can apply the conversion rules in Tables~\ref{Tab:OpLblRls} and~\ref{Tab:StrlRls} modularly (cf. Example~\ref{Exmpl:Lynd}).   
 
 \begin{Rem}    \label{Rem:ModCnv} 
 If $\lbl{L} \clrd \lbl{L}'$ and $\lbl{K} \clrd \lbl{K}'$, then $\ds{\lbl{L}}{\lbl{K}} \clrd \ds{\lbl{L}'}{\lbl{K}'}$.
\end{Rem}  

\begin{Prop}[Conversion] \label{Prop:Red} 
Every label  $\lbl{L}$ can be eventually converted   
 (by repeated applications of the  conversion rules in Tables~\ref{Tab:OpLblRls} and~\ref{Tab:StrlRls})   
to an equivalent  basic graph $\rd{\lbl{L}}$. 
\end{Prop}  

\subsection{Graph expansion} \label{subsec:Exprl} 
We now examine  graph expansion and its rule in our  calculus. 

\begin{Exmpl}  \label{Exmpl:Exprl} We now establish the inclusion 
$\RP \rp (\RQ \rs \RR) \, \sqeq \,  (\RP \rp \RQ) \rs \RR$. 
\Be 
\item As before, we  begin %
with the difference slice 
$\ds{ \RP \rp (\RQ \rs \RR) }{ (\RP \rp \RQ) \rs \RR }$.   
 \item We can convert it to a slice $\snt{S}'$ 
having  complemented slices as arc labels. 
With the following slices
\[ \SlT_1 :=  \inp \, \ndv_1 \, \strl{{\Ds \RR}}{\rar} \,  \nd{y}_1 \,  \inp, 
\SlT_2 :=  \inp \, \nd{x}_2 \, \strl{{\Ds \RP}}{\rar} \,  \ndw_2 \, 
	\strl{{\Ds \RQ}}{\rar} \,  \nd{y}_2 \,  \inp \mbox{ and} \]
\[ \xymatrix@C70pt{ 
\SlT_3 :=  
 \inp \, \ndu_3 \hs{\lvs pt}  \ar[r]^{ \hs{20 pt}  \cmpl{ \mbox{ \dashbox{\dsz}(64, 18)[]{$ \inp  \nd{u}_4 \, \strl{{\Ds \RQ}}{\rar}\,  \nd{z}_4  \inp $}}  } } &
 \ndv_3 \ar[r]^{ \cmpl{ \mbox{ \dashbox{\dsz}(64, 18)[]{$ \inp  \nd{z}_5 \, \strl{{\Ds \RR}}{\rar}\,  \nd{y}_5  \inp $}}  }  \hs{10 pt} } &
 \y_3 \, \inp },  \]    
 we have slice $\snt{S}'$ as follows: 
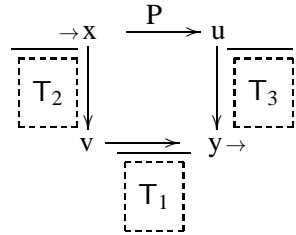
\[  \xymatrix@R30pt@C20pt{ 
	 \ar[d]_{{\Ds \cmpl{\mbox{ \dashbox{\dsz}(22, 26)[]{$\SlT_2$}} } \, }} \inp \, \x \hs{\mvs pt}  
	 \ar[r]^{{\Ds \, \RP}} & \, \ndu  \ar[d]^{ \, {\Ds \cmpl{\mbox{ \dashbox{\dsz}(22, 26)[]{$\SlT_3$}} } }} \\  
	 \ndv \, \ar[r]_{{\Ds \, \cmpl{{\Ds \mbox{ \dashbox{\dsz}(22, 26)[]{$\SlT_1$} } } }}}  &  \hs{\mvs pt} \y \, \inp 
	} \] 
\item This slice $\snt{S}'$ is not yet inconsistent. 
We can however expand it 
to a  graph $\snt{G}$ consisting of $2$ alternative slices  $\SlS_+$ and $\SlS_-$, 
respectively as follows:  
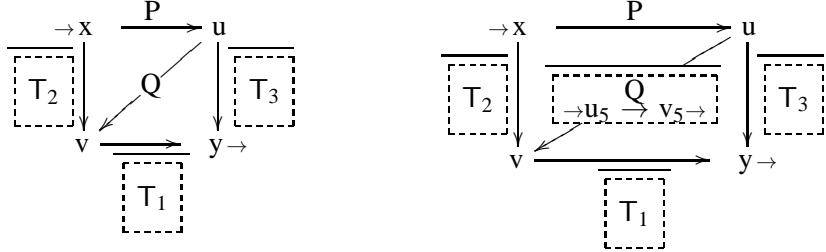
\[ \Ba{ccc} 
\xymatrix@R32pt@C22pt{ 
	 \ar[d]_{{\Ds \cmpl{\mbox{ \dashbox{\dsz}(22, 26)[]{$\SlT_2$}} } \, }} \inp \, \x \hs{\mvs pt}  
	 \ar[r]^{{\Ds \, \RP}} & \, \ndu  \ar[d]^{ \, {\Ds \cmpl{\mbox{ \dashbox{\dsz}(22, 26)[]{$\SlT_3$}} } }} 
 \ar[dl]|{{\Ds \RQ}}  \\ 
	\ndv \, \ar[r]_{{\Ds \cmpl{{\Ds \, \mbox{ \dashbox{\dsz}(22, 26)[]{$\SlT_1$} } } }}}  & \hs{\mvs pt} \y \, \inp }
& \hs{30 pt} & 
\xymatrix@R38pt@C58pt{ 
	 \ar[d]_{{\Ds \cmpl{\mbox{ \dashbox{\dsz}(22, 26)[]{$\SlT_2$}} } \, }} \inp \, \x \hs{\mvs pt}  
	 \ar[r]^{{\Ds \, \RP}} & \, \ndu  \ar[d]^{ \, {\Ds \cmpl{\mbox{ \dashbox{\dsz}(22, 26)[]{$\SlT_3$}} } }} 
 \ar[dl]|{{\Ds \cmpl{\mbox{ \dashbox{\dsz}(62, 18)[]{$ \inp  \nd{u}_5 \, 
	 \strl{{\Ds \RQ}}{\rar} \,  \ndv_5  \inp $}} \, }}}  \\ 
	\ndv \, \ar[r]_{{\Ds \, \cmpl{{\Ds \mbox{ \dashbox{\dsz}(22, 26)[]{$\SlT_1$} } } }}}  & \hs{\mvs pt} \y \, \inp  }
\Ea \] 
 \Ee 
Now, both $\snt{S}_+$ and $\snt{S}_-$  can be seen to be zero slices: 
slice $\snt{S}_+$ has the arcs $\x \, \RP \, \ndu$, $\ndu \, \RQ \, \ndv$ and 
$\x \, \cmpl{\SlT_2} \, \ndv$, while slice $\snt{S}_-$ has the arcs 
$\ndu \, \cmpl{\mbox{ \dashbox{\dsz}(58, 18)[]{$ \inp  \nd{u}_5 \, \strl{{\Ds \RQ}}{\rar} \,  \ndv_5  \inp $}}} 
\hs{3 pt} \ndv$, $\ndv \, \cmpl{\SlT_1} \, \y$ and 
$\ndu \, \cmpl{\SlT_3} \, \y$. 
Therefore,  we have established the inclusion 
$\{ \snt{S}_+ , \snt{S}_- \} \sqeq \, \btm$, whence also  
$\{ \snt{S}' \} \sqeq \, \btm$ and 
$\RP \rp (\RQ \rs \RR) \, \sqeq \, (\RP \rp \RQ) \rs \RR$. 
\end{Exmpl}  

	The expansion rule has an instance 
for slices $\snt{S}$ and $\snt{T}$  and pair of nodes $\Bop \ndu, \ndv \Eop$ 
of $ \snt{S}$, 
which  replaces  the single-slice graph $\{  \, \snt{S}  \, \}$ by the $2$-slice graph  
$\{ \, \gl{\snt{S}} {\ndu} {\ndv}  {\snt{T}} \, \} \, \cup \,   \{ \,   \snt{S} \, \ada \,  \ndu \,  \cmpl {\snt{T}} \, \ndv  \, \} $.    The \emph{ expansion rule} is as follows: 
\[ \Ba{lcl}  
 \mbox{{\sm ($\Exp$)}} \hs{10pt}
 \Ds \frac{ \{  \, \snt{S}  \, \}}
{ \{  \, \gl{\snt{S}} {\ndu} {\ndv}  {\snt{T}}     \, , \,  
  \snt{S} \, \ada \,  \ndu \,  \cmpl {\snt{T}} \, \ndv  \, \} } & 
\hs{5pt}  \hs{5pt} \Bop u, v \Eop \in {N_{\snt{S}}}^2 \hs{10pt} & 
\mbox{{\small replace $\snt{S}$  by 
$\gl{\snt{S}} {\ndu} {\ndv}  {\snt{T}}$} {\fns \&}  {\sm $\snt{S} \, \ada \,  \ndu \,  \cmpl {\snt{T}} \, \ndv$ }} 
\Ea  \] 

	We use $\ex$ for the \emph{immediate expansion} relation between graphs 
(e.g. $\{ \snt{S}' \} \ex \{ \SlS_+ , \SlS_- \}$  in Example~\ref{Exmpl:Exprl})
	and $\clrex$ for its reflexive-transitive closure: the \emph{eventual expansion} relation. 
A \emph{derivation} is a sequence $\lbl{L}, \snt{G}_0, \dots, \snt{G}_n$ of labels, such that, 
$\snt{G}_0, \dots, \snt{G}_n$ are graphs, $\lbl{L}$ eventually converts to $\snt{G}_0$ 
($\lbl{L} \, \clrd \, \snt{G}_0$) and, for each $i = 1, \dots, n$, 
$\snt{G}_{i-1}$ converts or expands to $\snt{G}_i$ ($\snt{G}_{i-1} \, ( \red \cup \ex) \, \snt{G}_i$). 
Call a derivation \emph{normal} iff applications of conversion rules precede  applications of expansions.\footnote{The 
preceding examples  use normal derivations: of the form $\lbl{L}  \, \clrd \, \snt{G} \, \clrex \, \snt{H}$.} 
We say that label $\lbl{L}$ \emph{derives} graph $\snt{H}$ (noted $\lbl{L} \, \Der \, \snt{H}$) iff 
there exists a derivation $\lbl{L}, \snt{G}_0, \dots, \snt{G}_n$ with $\snt{G}_n = \snt{H}$. 
Call a label  \emph{derivably zero} iff it derives some zero graph and  \emph{expansively zero} iff it eventually expands to some zero graph. 
 
	We have soundness and completeness of (normal) derivations. 

\begin{Thrm}[Correctness] \label{Thrm:Corr} 
Consider a label $\lbl{L}$. 
\Bd 
\item[(Sound)] If label $\lbl{L}$  is  derivably zero, then $\lbl{L}$  is null. 
\item[(Complete)] If label $\lbl{L}$  is  a null basic graph,  then $\lbl{L}$ is expansively zero. 
\Ed
\end{Thrm}  

	Soundness is not difficult to see. 
For establishing completeness, 
we introduce   (by mutual recursion)  
two measures of structural complexity: 
 \emph{rank} and \emph{set of embedded slices}, with the  aim of 
 providing an appropriate inductive measure. 
For  a relation name $r \in \RN$: $\rkl{ r } := 0$ and $\ESL{ r } := \emptyset$; 
for a basic slice $\SlT$:  $\rkl{ \cmpl{\snt{T}} } := \rks{ \snt{T} } + 1$ and 
$\ESL{ \cmpl{\snt{T}} } := \ESL{ \snt{T} } \cup \{  \snt{T} \} $.  
For a basic label $\lbl{L}$: 
$\rka{u \, \lbl{L} \, v} := \rkl{ \lbl{L} }$ and 
 $\ESA{u \lbl{L} \, v} := \ESL{ \lbl{L} }$.  
For a basic draft $\snt{D}$: $\rkd{\snt{D}} := \sum_{\snt{a} \in A_{\snt{D}}} \, \rka{\snt{a}}$ and 
for a basic sketch $\dg$: $\ESD{\dg} := \bigcup_{\snt{a} \in A_{\dg}} \, \ESA{\snt{a}}$.  
For a basic slice $\SlS$: $\rks{\SlS} := \rkd{\ul{\SlS}}$ and 
$\ESS{\SlS} := \ESD{\ul{\SlS}}$. 
Thus,  for a basic draft  
$\snt{D} = \snt{D}' \,  \ada \,  \ndu \,  \cmpl{\SlT} \, \ndv$, 
 with  $ \ndu \,  \cmpl{\SlT} \, \ndv \not \in A_{\snt{D}'}$,   
we will have 
$\rkd{ \snt{D} } = \rkd{ \snt{D}' } + \rkd{\SlT} + 1$ and 
 $ \ESD{\snt{D}} = \ESD{\snt{D}'} \cup \ESD{\snt{T}} \cup \{ \snt{T} \}$.  

	We now indicate how one can  establish completeness.  
Consider a basic graph $\snt{G}$  
that is not expansively zero. Then, it has a slice $\snt{S} \in \snt{G}$ that is not zero, such that, 
for every $\Bop \ndu, \ndv \Eop \in {N_{\snt{S}}}^2$ and  basic slice $\snt{T}$, 
$\{ \gl{\snt{S}} {\ndu} {\ndv}  {\snt{T}} \}$ or  $\{ \snt{S} \, \ada \, \ndu \,  \cmpl {\snt{T}} \, \ndv \}$ is not expansively zero.  
Thus, we can then obtain a family $\FRS$ of non-zero basic slices   
(with underlying drafts connected by morphisms), 
which is saturated by applications of the expansion rule.\footnote{
One may regard this as an analogue of Lindenbaum's Lemma:  extending a consistent theory 
to a maximally consistent one.}    
This family $\FRS$ can be used to obtain a co-limit sketch $\dg$,  
giving a 
 natural model  $\gC$ (cf.~\ref{subsec:Cnstr}),  
which discriminates satisfying assignments as morphisms to $\dg$: 
for a  basic draft $\snt{D}$ with $\ESD{\snt{D}} \incl  \ESD{\dg}$, we have 
 $\g : \snt{D} \rar \gC$ iff  $\g : \snt{D} \mor  \dg$ 
 (by induction on $\rkd{\snt{D}}$). 
Hence, we have 
a counter-model: $\betCn{\snt{G}} \supseteq \betCn{\snt{S}} \neq \ES$. 

We thus have a correct  calculus for null labels and for valid label inclusions. 
\Bd 
\item[{\sm ($\lbl{L}$)}] A label $\lbl{L}$  is  null iff it its basic form $\rd{\lbl{L}}$ is expansively zero. 
\item[{\sm ($\sqeq$)}] A label inclusion $\lbl{L} \, \sqeq \,  \lbl{K}$  is  valid iff  $\rd{\{ \ds{\lbl{L}}{\lbl{K}} \} }$ 
is expansively  zero. 
\Ed  

\section{Hypotheses} \label{sec:Hyp} 
We now extend the preceding ideas   to  handle inclusions as hypotheses, 
by  resorting  to difference slices. 

\begin{Exmpl}  \label{Exmpl:DerwoHyprl} Consider the assertion:    
``$\RP \rp \RR'  \rp \RQ \, \sqeq \,  \RP \rp \RR''  \rp \RQ$  follows from  $\RR' \, \sqeq \,  \RR''$''. 
 We reduce it  to deriving  
 $( \RP \rp \RR'  \rp \RQ ) \sa  \cmpl{ \RP \rp \RR''  \rp \RQ}  \sqeq \, \btm$ from 
 $ \RR'  \sa \cmpl{ \RR''}  \sqeq \,  \btm$. 

	The difference slice $\ds{\RR'}{ \RR'' }$  is equivalent to     
$\SlS'  := 
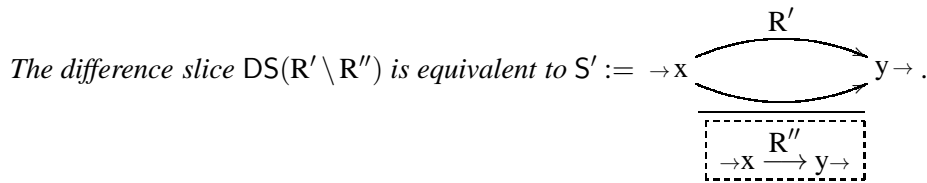
\xy
(0,0)*+{ \inp \, \x}="xS";
(30,0)*+{ \y \, \inp}="yS2";
{\aro{\RR' }@/_-1pc/"xS";"yS2"};
{\aru{{ \cmpl{{\Ds \mbox{ \dashbox{\dsz}(60, 22)[]{$\inp \x  \strl{{\Ds \RR'' }}{\lngrar} \y \inp$}}}} }}@/^-1pc/"xS";"yS2"}
\endxy$.
\Be 	
\item Begin with the graph $\{ \ds{\RP \rp \RR'  \rp \RQ}{ \RP \rp \RR''  \rp \RQ } \}$, with single slice 
$\snt{S}_0$ as follows:  
$$\snt{S}_0 \hs{\mvs pt}  :=  \hs{\mvs pt}  
	
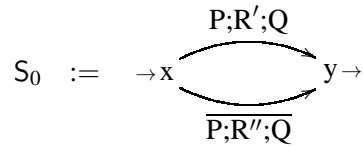
\xy
	(0,0)*+{ \inp \, \x}="xS";
	(25,0)*+{ \y \, \inp}="yS2";
	{\aro{ \RP \rp \RR' \rp \RQ }@/_-1pc/"xS";"yS2"};
	{\aru{{\cmpl{\RP \rp \RR'' \rp \RQ} }}@/^-1pc/"xS";"yS2"}
	\endxy $$

\item 
	Slice $\SlS_0$ is equivalent to the following slice $\snt{S}_1$: 
	\[ \xy
	(0,0)*+{ \inp \, \x \hs{\mvs pt} }="x";  (15,0)*++{\, \ndu}="u";
	(0,-15)*+{ \out \, \y \hs{\mvs pt} }="y";  (15,-15)*++{ \, \ndv}="v";
	{\aro{ \, \RP }@/_-0pc/"x";"u"}; {\aro{ \, \RQ }@/_-0pc/"v";"y"};
	{\ar_{{\Ds \cmpl{{\Ds \mbox{ \dashbox{\dsz}(116, 26)[]{
		$\inp \, \x'  \strl{{\Ds \RP}}{\rar} \, \ndu' \,  \strl{{\Ds \RR''}}{\rar} \, 
		\ndv' \,  \strl{{\Ds \RQ}}{\rar} \, \y' \, \inp$}}}} \, }}"x";"y"};  {\ar^{{\Ds \RR'}}"u";"v"}
	\endxy \] 

\item Now, expand graph $\{ \snt{S}_1 \}$  (with $\snt{T} := \inp \, \x \, \strl{{\Ds \RR'' }}{\rar} \,  \y \, \inp$), 
obtaining a graph  $\snt{H} = \{ \snt{S}_+ ,  \snt{S}_- \}$, where slices 
$\snt{S}_+ := \gl{\snt{S}_1} {\ndu} {\ndv} {\snt{T}}$ and 
	$\snt{S}_-  : = \snt{S}_1 \, \ada \,  \ndu \,  \cmpl{\snt{T}} \, \ndv$ are as follows: 
\[     \xy  (-60,-8)*+{ \snt{S}_+  \hs{\mvs pt} := \hs{\mvs pt} }="S"; 
	(0,0)*+{ \inp \, \x \hs{\mvs pt} }="x";  (15,0)*++{ \ndu}="u";
	(0,-15)*+{ \out \, \y \hs{\mvs pt} }="y";  (15,-15)*++{ \ndv}="v";
	{\aro{ \, \RP }@/_-0pc/"x";"u"}; {\aro{ \, \RQ }@/_-0pc/"v";"y"};
	{\ar_{{\Ds \cmpl{{\Ds \mbox{ \dashbox{\dsz}(116, 26)[]{
		$\inp \, \x'  \strl{{\Ds \RP}}{\rar} \, \ndu' \,  \strl{{\Ds \RR''}}{\rar} \, 
		\ndv' \,  \strl{{\Ds \RQ}}{\rar} \, \y' \, \inp$}}}} \, }}"x";"y"};  
		{\ar_{{\Ds \RR'}}@/_+1pc/"u";"v"};  {\ar^{{\Ds \RR''}}@/_-1pc/"u";"v"}; 
	\endxy \] 
\[     \xy  (-60,-8)*+{ \snt{S}_-  \hs{\mvs pt} := \hs{\mvs pt} }="S"; 
	(0,0)*+{ \inp \, \x \hs{\mvs pt} }="x";  (15,0)*++{ \ndu}="u";
	(0,-15)*+{ \out \, \y \hs{\mvs pt} }="y";  (15,-15)*++{ \ndv}="v";
	{\aro{ \, \RP }@/_-0pc/"x";"u"}; {\aro{ \, \RQ }@/_-0pc/"v";"y"};
	{\ar_{{\Ds \cmpl{{\Ds \mbox{ \dashbox{\dsz}(116, 26)[]{
		$\inp \, \x'  \strl{{\Ds \RP}}{\rar} \, \ndu' \,  \strl{{\Ds \RR''}}{\rar} \, 
		\ndv' \,  \strl{{\Ds \RQ}}{\rar} \, \y' \, \inp$}}}} \, }}"x";"y"};  
		{\ar_{{\Ds \RR'}}@/_+1pc/"u";"v"};  
		{\ar^{{\Ds \cmpl{ \mbox{ \dashbox{\dsz}(56, 22)[]{
		$\inp \, \x  \strl{{\Ds \RR''}}{\rar} \,  \y \, \inp$}}} }}@/_-1pc/"u";"v"}; 
	\endxy \] 

\Ee
Now, consider the graph $\snt{H} :=  \{ \snt{S}_+ ,  \snt{S}_- \}$. 
	\Bi 
	\item 
	Slice $\snt{S}_+ $ is zero (because  we have a morphism 
	$\hth$ from $\mbox{ \dashbox{\dsz}(100, 26)[]{
		$ \x'  \strl{{\Ds \RP}}{\rar} \, \ndu' \,  \strl{{\Ds \RR''}}{\rar} \, 
		\ndv' \,  \strl{{\Ds \RQ}}{\rar} \, \y' $}}$ to $\ul{\snt{S}_+}$, given by 
	$\x' \mapsto \x , \ndu' \mapsto \ndu  , \ndv' \mapsto \ndv , \y' \mapsto \y$). 
	\item 
	As for slice $\snt{S}_- $, we have a morphism 
	$\hth' : \ul{\snt{S}'} \mor \ul{\snt{S}_-}$, given by 
	$\x \mapsto \ndu , \y \mapsto \ndv$.  
	\Ei 
Thus, $\snt{H}$ has empty extension in any model where the hypothesis $\RR' \, \sqeq \, \RR''$ holds. 	 	 
\end{Exmpl}  

	Given a set $\SLI$ of   inclusions, we say that 
$\SLI$  \emph{holds} in  model $\gM$ 
(noted $\gM \models \SLI$)  iff every inclusion  in $\SLI$ holds in $\gM$. 
Now, we say that  inclusion $\lbl{L} \sqeq \lbl{K}$ \emph{follows from} set $\SLI$ of inclusions 
(noted $\SLI \models \lbl{L} \sqeq \lbl{K}$)
iff  
$\lbl{L} \sqeq \lbl{K}$  holds in every model $\gM$ where  $\SLI$  holds, i.e. 
 $\gM \models \lbl{L} \sqeq \lbl{K}$, whenever $\gM \models \SLI$.

	In Example~\ref{Exmpl:DerwoHyprl},  
we  have 
$\{ \snt{S}_0 \}  \eq \{  \snt{S}_- , \snt{S}_+ \}$, where  
$\snt{S}_+$ is a zero slice and one can erase  
slice $\snt{S}_- $. 

	Given a set $\SSH$ of slices, call a slice $ \snt{S}$ \emph{$\SSH$-erasable} iff $\, \Mor{\ul{\snt{S}}'} {\ul{\snt{S}}} \neq \emptyset \, $    for some $ \snt{S}' \in \SSH$. 
The rule for hypothesis  states that one can erase any $\SSH$-erasable slice. 
The  rule for hypothesis $\Hyp{\SSH}$ is as follows:  
\[ \Ba{lcl}  
 \mbox{{\sm ($\Hyp{\SSH}$)}}
  \hs{5pt}
 \Ds \frac{ \{ \snt{S} \}}
{ \{ \hs{3pt} \} } &\hs{10pt} & 
\mbox{ if slice $ \snt{S}$ is $\SSH$-erasable} 
\Ea  \]

	One can also  widen the goal to \emph{$\SSH$-zero} graphs, where each slice  is zero or 
$\SSH$-erasable. 
We have two versions of graph calculus with hypotheses. 
Given a set $\SSH$ of slices and a (basic) graph $\snt{G}$,  
we have two ways of establishing that $\snt{G} \, \sqeq \, \btm$ follows from the set of assumed inclusions 
$\ISS{\SSH} := \Setof{\snt{S}' \sqeq \btm}{\snt{S}' \in \SSH}$.  
\Bi 
\item 
Derive a zero graph by using the  rules   ($\Exp$) and   ($\Hyp{\SSH}$),   or 
\item  
 derive a $\SSH$-zero graph by using only the expansion rule  ($\Exp$). 
\Ei 
Both versions are sound and complete for a set $\SSH$ consisting of basic slices. 

\begin{Thrm}[Hypotheses] 
\label{Thrm:Hyp} 
Given a set $\SSH$ of basic slices and a basic graph $\snt{G}$, the following 3 assertions 
are equivalent. 
\Be 
\item  Inclusion $\snt{G} \, \sqeq \, \btm$ follows from 
$\ISS{\SSH} = \Setof{\snt{S}' \sqeq \btm}{\snt{S}' \in \SSH}$:  
$\ISS{\SSH} \models \snt{G} \sqeq \btm$. 
\item  From  $\snt{G}$ one can derive a zero graph by applications of 
 ($\Exp$) and   ($\Hyp{\SSH}$).  
\item  From $\snt{G}$ one can derive  a $\SSH$-zero graph by applications of  the rule  ($\Exp$).  
\Ee
\end{Thrm}  


\section{Conclusion} \label{sec:Cncl} 
We now present some concluding remarks about graph calculi for relational inclusions. 

	We have examined a sound and complete goal-oriented graphical calculus for inclusions:  
it reduces establishing a label inclusion to establishing that a graph constructed from it has empty extension. 
Relational terms, slices and graphs are labels and every label is equivalent to a basic graph and to a slice.\footnote{Also, 
any Boolean combination of   inclusions is equivalent to an inclusion 
$\lbl{L} \, \sqeq \,  \btm$~\cite{Mad_06}.}  



	Our goal-oriented calculus is simpler than some of  the available  graph relational 
calculi~\cite{FVVV_06,FVVV_07,FVVV_08,FVVV_09,FVVV_09',FVVV_10}. 
It is conceptually simpler as it proceeds  by eliminating relational operations  
and its rules 
require only the concept of (draft) morphism (rather than 
slice homomorphism -- a draft morphism that respects input and output nodes -- and 
graph cover~\cite{FVVV_08}). 
Also, it manipulates a single graph trying to convert it to a zero graph (rather than two graphs and comparing them~\cite{FVVV_10}). For instance, to establish directly the inclusion 
$\cnv{\Rr} \rp \, \cmpl{\Rr \rp \Rs} \, \sqeq \,  \cmpl{\Rs}$ (cf. Example~\ref{Exmpl:RAaxm}), 
one would have to apply the expansion rule.\footnote{The 
basic forms of these $2$ terms are single-slice graphs,   
with the following slices $\SlS$  and $\SlT$, respectively: 
$\xy 
(0,0)*+{ \inp \, \x}="xS"; (10,0)*+{ \z}="zS";  (46,0)*+{ \y \, \inp}="yS"; 
{\ar_{{\Ds \, \Rr}}@/_-0pc/"zS";"xS"}; 
{\ar^{{\Ds \cmpl{ \mbox{ \dashbox{\dsz}(66, 26)[]{$\inp \x \, \strld{ \Rr}{\rar}  \z \, 
	 \strld{ \Rs}{\rar} \, \y  \, \inp $}} }}}@/_-0pc/"zS";"yS"}
\endxy$ and 
$\xy 
(0,0)*+{ \inp \, \x}="xT";  (28,0)*+{ \y \, \inp}="yT"; 
{\ar^{{\Ds \cmpl{ \mbox{ \dashbox{\dsz}(46, 26)[]{$\inp \x \, \strld{ \Rs}{\rar} \, \y  \, \inp $}} }}}@/_-0pc/"xT";"yT"}
\endxy$. 
There is no homomorphism from $\SlT$  to $\SlS$. We can, however, expand graph $\{ \SlS \}$ to a graph 
$\snt{G} := \{ \SlS_- , \SlS_+ \}$, with slices $\SlS_-$  and $\SlS_+$, respectively,  as follows: 
$\xy 
(0,0)*+{ \inp \, \x}="xS"; (10,0)*+{ \z}="zS";  (46,0)*+{ \y \, \inp}="yS"; 
{\ar_{{\Ds \, \Rr}}@/_-0pc/"zS";"xS"}; 
{\ar^{{\Ds \cmpl{ \mbox{ \dashbox{\dsz}(66, 26)[]{$\inp \x \, \strld{ \Rr}{\rar}  \z \, 
	 \strld{ \Rs}{\rar} \, \y  \, \inp $}} }}}@/_-0pc/"zS";"yS"}; 
{\ar_{{\Ds \, \Rs}}@/_+1pc/"xS";"yS"} 
\endxy$ and 
$\xy 
(0,0)*+{ \inp \, \x}="xS"; (10,0)*+{ \z}="zS";  (46,0)*+{ \y \, \inp}="yS"; 
{\ar_{{\Ds \, \Rr}}@/_-0pc/"zS";"xS"}; 
{\ar^{{\Ds \cmpl{ \mbox{ \dashbox{\dsz}(66, 26)[]{$\inp \x \, \strld{ \Rr}{\rar}  \z \, 
	 \strld{ \Rs}{\rar} \, \y  \, \inp $}} }}}@/_-0pc/"zS";"yS"}; 
{\ar_{{\Ds \cmpl{ \mbox{ \dashbox{\dsz}(46, 26)[]{$\inp \x \, \strld{ \Rs}{\rar} \, \y  \, \inp $}} }}}@/_+1pc/"xS";"yS"}
\endxy$. 
Now, slice  $\SlS_-$  is a zero slice 
(cf. slice  $\snt{S}_4$ in Example~\ref{Exmpl:RAaxm} in Section~\ref{sec:BsIds}) and 
we have a homomorphism from $\SlT$  to $\SlS_+$.}  
In fact, whenever there is a slice homomorphism from $\SlT$  to $\SlS$, the difference slice 
$\ds{\SlS}{ \SlT }$ is a zero slice. 
	 
	Also, the treatment of hypotheses 
is much simpler than in the usual calculi,  
as it resorts to erasing (rather than gluing) slices. 
The assertion in Example~\ref{Exmpl:DerwoHyprl} can be established directly without the expansion rule 
(by means of the gluing rule for hypotheses). 
On the other hand,  an assertion like 
``$\Rr  \, \sqeq \, \Rs$ follows from $\Rr \sa \cmpl{\Rs} \, \sqeq \, \btm$'', which is trivial in our approach, 
will require using the expansion rule in the direct approach.\footnote{With the slices 
$\SlS := \inp \, \strld{\Rr}{\rar} \, \y  \, \inp$ and $\SlT := \inp \, \strld{\Rs}{\rar} \, \y  \, \inp$, 
as there is no homomorphism from $\SlT$  to $\SlS$, we have to expand  graph $\{ \SlS \}$ to a graph 
$\snt{G} := \{ \SlS_- , \SlS_+ \}$, with slices $\SlS_-$  and $\SlS_+$, respectively,  as follows: 
$
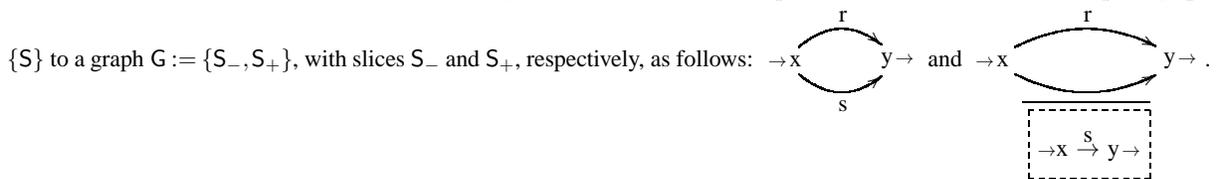
\xy 
(0,0)*+{ \inp \, \x}="x"; (15,0)*+{ \y \, \inp}="y";  
{\ar^{{\Ds \, \Rr}}@/_-1pc/"x";"y"}; 
{\ar_{{\Ds \, \Rs}}@/_+1pc/"x";"y"} 
\endxy$ and 
$\xy 
(0,0)*+{ \inp \, \x}="x"; (25,0)*+{ \y \, \inp}="y"; 
{\ar^{{\Ds \, \Rr}}@/_-1pc/"x";"y"}; 
{\ar_{{\Ds \cmpl{ \mbox{ \dashbox{\dsz}(46, 26)[]{$\inp \x \, \strld{ \Rs}{\rar} \, \y  \, \inp $}} }}}@/_+1pc/"x";"y"}
\endxy$. 
Now, we have a homomorphism from $\SlT$  to $\SlS_-$ and slice  $\SlS_+$  can be erased  
(as we have a homomorphism from $\SlS$  to $\SlS_+$).} 

	Moreover, the idea of labels with embedded  slices or graphs  is rather  powerful. 
Comparing with other graph calculi, we conjecture that there is not much gain or loss in complexity order, 
its main advantages are  on the  conceptual side: simpler concepts and goal orientation. 


\nocite{*}
\bibliographystyle{eptcs}
\bibliography{generic}

\end{document}